\documentclass[12pt,letterpaper]{article}

\usepackage[includeheadfoot,
            marginratio={1:1,2:3}, 
            width=412pt, 
            height=688pt,]{geometry}

\usepackage[utf8]{inputenc}
\usepackage{amsmath}
\usepackage{amsfonts}
\usepackage{amssymb}
\usepackage{graphicx}
\usepackage{booktabs}
\usepackage{empheq}
\usepackage{paralist}
\usepackage{cite}
\usepackage[hidelinks]{hyperref}
\usepackage{xcolor}
\usepackage{tensor}
\usepackage{tikz}
\usepackage{adjustbox}
\usepackage{subcaption}

\newcommand{\nc}{\newcommand}
\nc{\lb}{\llbracket}
\nc{\rb}{\rrbracket}
\nc{\gl}{\llbracket}
\nc{\gr}{\rrbracket}
\nc{\del}{\partial}
\nc{\tri}{\hspace{-3.5pt}\vartriangle\hspace{-3.5pt}}
\nc{\blacktri}{\blacktriangle}
\nc{\eq}[1]{\begin{equation}
                     \begin{split} #1 \end{split}
                     \end{equation}}
\nc{\ul}{\underline}
\nc{\ov}{\overline}
\nc{\fa}{\hat}
\nc{\fb}{\MakeUppercase}
\nc{\fc}{\tilde }
\nc{\Lie}{{\cal L}} 
\nc{\lambdabar}{{\mkern0.75mu\mathchar '26\mkern -9.75mu\lambda}}

\allowdisplaybreaks[2]
\numberwithin{equation}{section}

\begin{document}

\vspace*{-1.5cm}
\begin{flushright}
  {\small
  MPP-2020-165\\
  }
\end{flushright}

\vspace{1.5cm}
\begin{center}
{\LARGE
Small Flux Superpotentials for Type IIB \\[0.2cm]
Flux Vacua Close to a Conifold
}
\vspace{0.4cm}
\end{center}

\vspace{0.35cm}
\begin{center}
 Rafael \'Alvarez-Garc\'ia$^{1,2}$,
 Ralph Blumenhagen$^{1}$, 
 Max Brinkmann$^{1}$, \\[0.1cm]
 Lorenz Schlechter$^{1}$
\end{center}

\vspace{0.1cm}
\begin{center} 
\emph{
$^{1}$
Max-Planck-Institut f\"ur Physik (Werner-Heisenberg-Institut), \\ 
   F\"ohringer Ring 6,  80805 M\"unchen, Germany } 
   \\[0.1cm] 
\vspace{0.25cm} 
\emph{$^{2}$ Ludwig-Maximilians-Universit{\"a}t M\"unchen, Fakult{\"a}t f{\"u}r Physik,\\ 
Theresienstr.~37, 80333 M\"unchen, Germany}\\
\vspace{0.2cm}

\vspace{0.3cm}
\end{center}

\vspace{0.5cm}


\begin{abstract}
We generalize the recently proposed mechanism by 
Demirtas, Kim, McAllister and Moritz\cite{Demirtas:2019sip} for  the explicit
construction of type IIB flux vacua with $|W_0|\ll 1$ to the region
close to the conifold locus in the complex structure moduli space.
For that purpose tools are developed to determine the periods
and the resulting prepotential close to such a codimension one
locus with all the remaining moduli still in the large complex
structure regime. As a proof of principle we present a working example
for the Calabi-Yau manifold $\mathbb{P}_{1,1,2,8,12}[24]$. 
\end{abstract}

\clearpage

\tableofcontents


\section{Introduction}
\label{sec:intro}

In view of the recent swampland conjectures, one has to revisit the
standard constructions of dS vacua in string theory. The most
recognized  approach is the mechanism of  KKLT \cite{Kachru:2003aw} in
which an initial non-perturbative AdS vacuum is uplifted by  anti
D3-branes placed at the tip of a strongly warped throat. This
construction has been scrutinized from various points of view. First
the dS uplift mechanism was questioned, namely whether a
$\overline{D3}$-brane at the tip of a warped throat is really a stable
configuration (see \cite{Danielsson:2018ztv} for a review). Moreover,
it has been questioned whether the  4D description of the KKLT AdS
minimum does really uplift to a full 10D  solution of string theory
\cite{Moritz:2017xto,Kallosh:2018wme,Kallosh:2018psh,Gautason:2018gln,Hamada:2018qef,Hamada:2019ack,Carta:2019rhx,Gautason:2019jwq,Bena:2019mte,Blumenhagen:2020dea}. Another
important point is whether the effective action that is presumed to
describe the strongly warped regime is really under control. Based on
earlier work \cite{Douglas:2007tu},  this question has been addressed 
recently \cite{Bena:2018fqc,Blumenhagen:2019qcg,Bena:2019sxm,Dudas:2019pls,Randall:2019ent}.

However, there is one even more basic assumption in the KKLT construction, which is that the three-form flux induced no-scale potential admits (Minkowski) vacua with an exponentially small value of $W_0$\footnote{
An argument has been made that large $W_0$ values can directly produce a supergravity potential with a dS minimum \cite{Linde:2020mdk}. However, it is not clear that this supergravity solution uplifts to a true solution of string theory.
}. There exist statistical arguments (see \cite{Douglas:2006es} for a
review) that this should be the case. However, based on an older
proposal \cite{Giryavets:2003vd,Denef:2004dm}, only very recently
Demirtas, Kim, McAllister and Moritz (DKMM)  \cite{Demirtas:2019sip}
formulated a concrete two-step mechanism for the explicit construction
of such vacua. Working  in the {\it large complex structure regime}
of a Calabi-Yau (CY) manifold one first considers only the leading order terms in the periods and
dials the fluxes such that one gets a supersymmetric minimum with
$W_0=0$. In fact this leaves at least one complex structure modulus
unstabilized. Taking in the second step also the non-perturbative
corrections to the periods into account, this final modulus also gets
frozen in a race-track manner and generically gives an exponentially
small $|W_0|\ll 1$ as a potential starting point for KKLT. However, for the final uplift one actually needs a similar
controllable mechanism  close to a conifold point in the complex
structure moduli space  where large warping can occur \cite{Giddings:2001yu}.

It is the objective of this paper to analyze whether such a
generalization of the DKMM mechanism can indeed be found. For that
purpose, one first needs to know the explicit form of the periods
close to a conifold locus in the complex structure moduli
space. The best studied example is the quintic for which the periods
in this regime have been determined 
explicitly \cite{Candelas:1990rm,Curio:2000sc,Huang:2006hq}
and for which there were already studies of moduli stabilization \cite{Bizet:2016paj,Blumenhagen:2016bfp}.
As we will describe, in this case finding models with $|W_0|\ll 1$ appears to be a number theoretic problem, i.e. there is no generic algorithm
where at leading order one starts with $W_0=0$ and then subleading
instanton corrections provide the exponentially small corrections.
The obstruction here is that besides the conifold modulus there are no other complex structure moduli
that can still take values in their large complex structure regime.

Thus the generalization of the DKMM mechanism requires some remaining complex
structure moduli to still be in their large complex structure regime. It
turns out that this regime that lies at the tangency of the conifold 
and the large complex structure locus has poorly been studied so far. While the
periods in the large complex structure (LCS) regime and deep in the
non-geometric regimes are well studied, an
equally satisfying method for points close to the conifold for more
than one modulus models is still lacking. Therefore, a large portion
of this paper deals with the development of such methods to compute the
relevant periods. 

Describing this in more detail, in the one modulus case a
possibility is to compute local solutions of the Picard-Fuchs (PF) equations and
transforming them into the symplectic basis analytically continued
from the LCS region. This approach was used e.g. in
\cite{Curio:2000sc,Huang:2006hq,Alim:2012ss,Bizet:2016paj,Joshi:2019nzi,Blumenhagen:2018nts}. While
it is in principle applicable in the multi-modulus case,
the computations become very tedious and have to be performed for
every point one is interested in separately. Another approach is to
work with the $\ov{\omega}$ periods of
\cite{Candelas:1990rm,Candelas:1993dm,Berglund:1993ax,Candelas:1994hw}
determined deep in the LCS/non-geometric phases. Unfortunately these converge badly at the
conifold. Nevertheless, it is possible to extract the periods using
very high orders, as is done for example in
\cite{Conlon:2004ds}. Methods based on gauged linear sigma models (GLSM)
\cite{Jockers:2012dk} face the same problem of slow convergence at the
phase boundaries. Recently, the Mellin-Barnes representations arising
from the GLSM were used in a  recursive construction, resulting in
infinite sum expressions for the entries of the transition 
matrix \cite{Scheidegger:2016ysn,Knapp:2016rec}.

Instead of a singular approach, we will apply a combination of various
methods. We will still try to find the transition matrix from a local
solution to the symplectic basis. To improve the numerics, monodromy
considerations as well as a symplectic form on the solution space
developed in \cite{Masuda:1998eh}, where it was applied to the
Seiberg-Witten point of the $\mathbb{P}_{1,1,2,2,6}[12]$ model, are
used. In \cite{Alim:2012ss} the same method was applied to the
$\mathbb{P}_2$-fibration phase of the $\mathbb{P}_{1,1,1,6,9}[18]$
CY. We then obtain an analytic solution for $\mathbb{P}_1$-fibrations
and compare it to the numerical results. We find good agreement
between the two methods.

This paper is structured as follows. Before we delve into the problem
of moduli stabilization, in section \ref{sec:periods_at_special_points_in_moduli_space} we start with the 
mathematically rather involved description of a
systematic way to compute the periods of a multi-parameter Calabi-Yau
manifold close to the point of tangency of the conifold with the
LCS regime analytically. The less mathematically inclined reader may
essentially skip this section after noticing
the result for the prepotential shown in equation  \eqref{prepotimp}, 
which we will use for our working example.
In section \ref{sec:quest_for_W0} we present a three-step mechanism to generate small $W_0$ vacua 
with all complex structure moduli as well as the axio-dilaton stabilized.  
Here we first discuss the example of the quintic and the appearing
obstructions to the formulation of such a mechanism for too simple models,
and then show that more involved examples behave much better.
Finally, we demonstrate the mechanism by constructing an
explicit example for the $\mathbb{P}_{1,1,2,8,12}[24]$ Calabi-Yau.

\vspace{0.5cm}
\noindent
{\it Note added:} While finishing this work we became aware of an upcoming
  paper \cite{Demirtas:2020ffz} by  Demirtas, Kim, McAllister and Moritz which approaches the same question.

\section{Periods at special points in moduli space}
\label{sec:periods_at_special_points_in_moduli_space}

In this section  we present the tools that we employed in order to compute
 a symplectic basis of periods close to the conifold locus with
the remaining moduli in their large complex structure regime.
This involves quite some mathematical machinery. For readers not
such interested in the technical details, we note that the main result
is the prepotential shown in equation \eqref{prepotimp}. This will be
employed in the upcoming section on moduli stabilization.

We will mainly focus on hypersurfaces (or complete intersections thereof) in weighted projective spaces $\mathbb{P}^n(\vec{w})$ defined by the zero locus of $m$ quasihomogeneous polynomials $P_i$ of degree $d_i$ satisfying 
\begin{equation}
\sum_{i=1}^m d_i=\sum_{j=1}^{n+1} w_j\,.
\end{equation}
These can be described by means of toric geometry through an n-dimensional convex integral reflexive polyhedron.
For the detailed construction and the topological properties of the resulting varieties see \cite{Hosono:1993qy} and references therein. 

The integral points $\nu_i$ of the polyhedron are embedded into $\mathbb{R}^{n+1}$ as $\overline{\nu}=(1,\nu_i)$. These are not linearly independent and their dependencies can be described by a lattice 
\begin{equation}
	L = \left\{(l_0,\dots,l_p)\in \mathbb{Z}^{p+1}\, \middle|\, \sum_{i=0}^pl_i\,\overline{\nu_i}=0\right\}\,,
\label{eq:Mori_lattice}
\end{equation}
whose basis $\{l_{i}\}$ can be chosen to be the one for the Mori cone
(cf. \cite{Hosono:1993qy}). This basis represents the charge matrix of
the associated GLSM and directly relates to the 
Gel'fand-Kapranov-Zelevinski (GKZ) hypergeometric system \cite{Gel'fand1989}
\begin{align}
    \mathcal{D}_l &= \prod_{l_i>0}\left({\frac{\partial}{\partial_{a_i}}}\right)^{l_i}-\prod_{l_i<0}\left({\frac{\partial}{\partial_{a_i}}}\right)^{-l_i}\,,\qquad l\in \{l_{i}\}\,,
\label{eq:GKZ}\\
    \mathcal{Z}_{j} &= \sum_{i= 0}^{n} \overline{\nu}_{i,j} \,a_{i}\, \frac{\partial}{\partial a_{i}} - \beta_{j}\,, \qquad\qquad j\in\{0,1,...,n\}\,,
\label{eq:GKZ_constraint}
\end{align}
which in turn  annihilates the period integrals \eqref{eq:period_integrals} we shall define in a moment. 
The $a_i$ are coordinates of an affine $\mathbb{C}^{p+1}$, which is larger than the physical complex structure moduli space. $\beta$ is a constant vector with $\beta_0=-1$ and $\beta_j=0$ for $j\neq 0$. The relevant coordinates around the LCS point are also given by the Mori cone basis as
\begin{equation}
	x_k=(-1)^{l_0^{(k)}}a_0^{l_0^{(k)}}\dots a_s^{l_s^{(k)}}
\label{eq:LCS_coordinates}
\end{equation}
where s is the number of vertices in the polyhedron. 
These are chosen such that any function of them is automatically annihilated by the $\mathcal{Z}_{j}$ of \eqref{eq:GKZ_constraint}.

In the appendix of \cite{Hosono:1993qy} the $l^{(i)}$-vectors and resulting Picard-Fuchs (PF) operators
$\mathcal{D}_i$, $i=1, \dots, h^{2,1}$ are listed for many examples. 
Note that for some models the PF operators obtained from the GKZ system \eqref{eq:GKZ} are not the complete PF system, requiring an extension which was also worked out in \cite{Hosono:1993qy}. 
For the example we will consider this is not necessary.
 
The periods we are ultimately interested in are defined by integrals over the unique holomorphic $(3,0)$-form as
\begin{equation}
\Pi(x)^{\alpha}= \int_{\gamma^\alpha} \Omega(x)\, ,\quad \gamma^{\alpha}\in H_3(X,\mathbb{Z})\, ,\quad\alpha=0,\dots,2 h^{2,1}+1\,.
\label{eq:period_integrals}
\end{equation}
Here X denotes the CY we are investigating.
The periods are annihilated by the PF operators.

\subsection{Computing the local periods}
\label{subsec:computing_local_periods}

For cleaner notation we use multi-indices 
$i=\{i_l\}_{l\in\{1,...,h^{2,1}\}}$, $\alpha_{j,k}=\{\alpha_{j,k,l}\}_l$, 
and $\beta_{j}=\{\beta_{j,l}\}_l$ in the following,
and employ the shorthand notations $x^i=\prod_{l=1}^{h^{2,1}} (x_l)^{i_l}$ and 
$(\log{x})^{\alpha_{j,k}} = \prod_{l=1}^{h^{2,1}}(\log{x_{l}})^{\alpha_{j,k,l}}$.
As the periods are annihilated by the PF operators, local solutions can be obtained by inserting the ansatz
\begin{equation}
	\omega_j=\sum_k\sum_{i} c_{i,j,k} \, x^{i+\beta_{j}}\, (\log{x})^{\alpha_{j,k}}\;,
\label{eq:ansatz}
\end{equation}
into the equations
\begin{equation}
\label{pf}
	\mathcal{D}_i \, \omega_j=0\,,
\end{equation}
expressed in coordinates centered around the point of interest. The sum over $k$ runs from $1$ to $4^{h^{2,1}}$ in order to capture terms with $h^{2,1}$ individual $\log$ factors, each with powers ranging from $0$ to $3$.
The sum over $i$ runs over the integer lattice $i_l\in\{0,...,m\}$, $l\in\{0,...,h^{2,1}\}$ with $m$ the order up to which we are computing. 
The multi-indices $\alpha_{j,k}$ and $\beta_{j}$ are made up of positive constants and can be fixed as follows.

The fundamental period $\omega_0$ with $\alpha_{0,k}=\beta_0=0$ is always present. 
Moreover, for each distinct $\beta_j$ there is one period with $\alpha_{j,k}=0$ $\forall k$, 
i.e. a pure power series. Thus, one makes the ansatz
\begin{equation}
\omega_j=\sum_{i} c_{i,j} \, x^{i+\beta_j}\;.
\end{equation}
The allowed $\beta_{j}$ are fixed by demanding the vanishing of the coefficients of $c_{0,j}$ in the constraints $\mathcal{D}_i w_j=0$.

The remaining periods have $ \alpha_{j,k,l} \in \{0,1,2,3\}$. 
For one parameter models the $\alpha_{j,k}$ associated with
each distinct $\beta_{j}$ range from 0 to the degeneracy of 
$\beta_{j}$ as a solution to the indicial equations. For multivariate
cases it is no longer clear how to count the degeneracies, as there is
more than one PF operator.

In that case the $\alpha_{j,k}$ can be determined by making the most
general ansatz as in (\ref{eq:ansatz}). 
To reduce the computation time one can restrict the $\alpha_{j,k}$ further by using the following
observation. The logarithmic structure of the periods with $\beta_j=0$
is completely fixed by the logarithmic structure at the LCS
point. Performing on the LCS periods the coordinate transformation to
coordinates centered around the point one is interested in and
expanding them in a series around the origin in the new coordinates,
gives exactly the needed structure. 
For the resulting expressions in a three-parameter example see appendices 
\ref{subsec:local_LCS_periods} and \ref{subsec:local_coni_periods}.

In \cite{Hosono:1994ax} a solution of \eqref{pf} around the LCS point in terms of the Mori-cone basis $\{l_{i}\}$ and the triple intersection numbers $K_{ijk}$ was given for any CICY. The fundamental period is
\eq{
	\omega_0 = \sum_{\substack{n_i=0\\i=1,\dots,h^{2,1}}}^\infty
        \left(\prod_{i=1}^{h^{2,1}}
          x_{i}^{n_i+\rho_i}\right) &\frac{\Gamma\left[1-\sum_{k=1}^{h^{2,1}}
            l_k^{(0)}(n_k+\rho_k)\right]}{\Gamma\left[1-\sum_{k=1}^{h^{2,1}}
            l_k^{(0)}\rho_k\right]} \\
&\phantom{aaaaaaa}\cdot\prod_{j=1}^{p}\frac{\Gamma\left[1-\sum_{k=1}^{h^{2,1}}
    l_k^{(j)}\rho_k\right]}{\Gamma\left[1-\sum_{k=1}^{h^{2,1}}
    l_k^{(j)}(n_k+\rho_k)\right]}\;.
}
In this expression $\rho_i\in\mathbb{R}$ are introduced, extending the summation variables $n_i$. Defining then the derivative operators with respect to them,
\begin{equation}
\begin{aligned}
    D_{1,i} &= \frac{1}{2\pi i} \partial_{\rho_{i}}\,,\\
    D_{2,i} &= \frac{1}{2} \frac{K_{ijk}}{(2\pi i)^{2}} \partial_{\rho_{j}}\partial_{\rho_{k}}\,,\\
    D_{3} &= -\frac{1}{6} \frac{K_{ijk}}{(2\pi i)^{3}} \partial_{\rho_{i}}\partial_{\rho_{j}}\partial_{\rho_{k}}\,,
\end{aligned}
\end{equation}
the period vector $\omega$ is computed at the natural indices $\rho_i=0$ as
\begin{equation}
\label{hyperbasis}
    \omega =
    \left.
    \begin{pmatrix}
    \omega_{0}\\
    D_{1,i}\, \omega_{0}\\
    D_{2,i}\, \omega_{0}\\
    D_{3}\, \omega_{0}
    \end{pmatrix}
    \right\rvert_{\rho_{i} = 0}
\end{equation}
with $i = 1, \dots, h^{2,1}$.

\subsection{An integral symplectic basis}
\label{subsec:integer_symplectic_basis}

The local periods need to be combined into an integer symplectic basis $\Pi$. The two bases are related by a linear transformation
\begin{equation}
	\Pi=m\cdot \omega\;.
\end{equation}
At the LCS and orbifold point the $(h^{2,1}+2)\times (h^{2,1}+2)$
matrix $m$ can be determined purely based on monodromy arguments. For
the conifold point the monodromies constrain the form of the
transition matrix but do not completely fix it. Moreover, the
monodromy calculations can become very involved for models with many
moduli, especially when the moduli space needs to be blown up and
monodromies around exceptional divisors are needed. 

In principle one could simply use a general ansatz for $m$ and
numerically determine it in the overlap of the regions of convergence
of $\Pi$ and $\omega$. But this turns out to be numerically
unstable. For one parameter models it is possible to go to very high
order and obtain reasonable results, but already for two parameter
models the convergence of the values is extremely slow. Thus, a
systematic method to constrain the transition matrix $m$ is needed.

An alternative to monodromy arguments is the use of the symplectic
form on the solution space of the PF operators. This symplectic form
was first introduced in \cite{Masuda:1998eh}, and in
\cite{Alim:2012ss} the same method was applied to the
$\mathbb{P}_2$-fibration phase of the $\mathbb{P}_{1,1,1,6,9}[18]$
CY. Since the space of periods and the space of solutions to the PF
system can be identified, one can represent the symplectic form
pairing the periods as a bilinear differential operator acting on the
space of PF solutions.

The symplectic form $Q$ is then given by
\begin{equation}
	Q(f_1,f_2)=\sum_{k,l}Q_{k,l}(x)\, D_k(f_1)\wedge D_l(f_2)\;,
\label{eq:symplectic_form}
\end{equation}
where $Q_{k,l}(x)$ are functions of the coordinates and $k$ and $l$ range over a basis of the ring of differential operators where $\mathcal{D}_j=0$. One imposes the conditions
\begin{equation}
    \frac{\partial}{\partial x_{i}} Q = 0\,,
\label{eq:constancy_symplectic_form}
\end{equation}
enforcing constancy over the moduli space, which leads to a system of coupled linear differential equations for the $Q_{k,l}(x)$. This system allows for the determination of the symplectic form up to an overall normalization $\eta$.

The symplectic form in turn enables us to systematically constrain the transition matrix $m$ by demanding that the resulting periods have the desired intersections. This does not fix it completely as there are combinations of the periods that leave the intersection matrix invariant. For example, given a certain $\alpha$-cycle that has vanishing intersections with all the other cycles except for its $\beta$-cycle companion, we could add multiples of the $\alpha$-cycle to the $\beta$-cycle without changing the value of the intersection. Therefore we still need to perform the numerical matching as a final step.

However, for our later construction it will be important that the coefficients of certain moduli in the prepotential are rational numbers. As it is not possible to prove this property by a numerical approach we describe in a later section a way to analytically compute the coefficients, thereby proving that they are indeed rational numbers.

\subsection{A (not so) special example \texorpdfstring{$\mathbb{P}_{1,1,2,8,12}[24]$}{P(1,1,2,8,12)[24]}}
\label{CY-P112812}

As a concrete geometry on which to explore the above constructions we choose (the mirror dual to) the three-parameter ($h^{1,1} = 3$, $h^{2,1} = 243$) CY $\mathbb{P}_{1,1,2,8,12}[24]$, given by the vanishing locus of the defining polynomial
\begin{equation}
\begin{aligned}
    P(x) &= \frac{1}{24}z_{1}^{24} + \frac{1}{24}z_{2}^{24} + \frac{1}{12}z_{3}^{12} + \frac{1}{3}z_{4}^{3} + \frac{1}{2}z_{5}^{2}\\
    &\phantom{=}- \psi_{0} z_{1}z_{2}z_{3}z_{4}z_{5} - \frac{1}{6}\psi_{6}(z_{1}z_{2}z_{3})^{6} - \frac{1}{12}\psi_{7}(z_{1}z_{2})^{12}
\end{aligned}
\end{equation}
in $W\mathbb{CP}^{4}_{1,1,2,8,12}$. The hypersurface can be seen as a fibration of a K3 surface, and contains a singular $\mathbb{Z}_{2}$ curve $C$, with an exceptional divisor corresponding to a $C \times \mathbb{P}^{1}$ surface, and an exceptional $\mathbb{Z}_{4}$ point, which is blown up to a Hirzebruch surface $\Sigma_{2}$. It has been studied e.g. in \cite{Hosono:1993qy,Kachru:1995fv,Kachru:1995wm,Klemm:1995tj,Suzuki:1997wn}.

The extended set of integral points in its toric geometric construction is given in \cite{Hosono:1993qy} as
\begin{equation}
\begin{aligned}
    \nu_{0} &= (0,0,0,0),\quad \nu_{1} = (1,0,0,0),\quad \nu_{2} = (0,1,0,0),\quad \nu_{3} = (0,0,1,0),\\ \nu_{4} &= (0,0,0,1),\quad \nu_{5} = (-12,-8,-2,-1),\quad \nu_{6} = (-3,-2,0,0),\\ \nu_{7} &= (-6,-4,-1,0),
\end{aligned}
\end{equation}
and after the embedding into $\mathbb{R}^{5}$ leads to a lattice of dependencies (\ref{eq:Mori_lattice}) from which we extract the Mori cone basis\footnote{Notice a typing error in \cite{Hosono:1993qy} affecting $l_{1}$ and $\mathcal{D}_{3}$.}
\begin{equation}
\begin{aligned}
    l_{1} &= (-6,3,2,0,0,0,1,0),\\
    l_{2} &= (0,0,0,0,1,1,0,-2),\\
    l_{3} &= (0,0,0,1,0,0,-2,1).
\end{aligned}
\end{equation}
These give through \eqref{eq:LCS_coordinates} the appropriate LCS coordinates
\begin{equation}
\begin{aligned}
    x &= \frac{a_{1}^{3}a_{2}^{2}a_{6}}{a_{0}^{6}} = -\frac{\psi_{6}}{2^{16}3^{6}\psi_{0}^{6}},\\
    y &= \frac{a_{4}a_{5}}{a_{7}^{2}} = \frac{2^{3}3}{\psi_{7}^{2}},\\
    z &= \frac{a_{3}a_{7}}{a_{6}^{2}} = -\frac{\psi_{7}}{2^{2}\psi_{6}^{2}}.
\end{aligned}
\end{equation}
The Picard-Fuchs operators in these coordinates read\footnotemark[\value{footnote}]
\begin{equation}
\begin{aligned}
    \mathcal{D}_{1} &= \Theta_{x} (\Theta_{x} - 2\Theta_{z}) - 12x(6\Theta_{x} + 5)(6\Theta_{x} + 1),\\
    \mathcal{D}_{2} &= \Theta_{y}^{2} - y(2\Theta_{y} - \Theta_{z} + 1)(2\Theta_{y} - \Theta_{z}),\\
    \mathcal{D}_{3} &= \Theta_{z} (\Theta_{z} - 2\Theta_{y}) - z(2\Theta_{z} - \Theta_{x} + 1)(2\Theta_{z} - \Theta_{x}),
\end{aligned}
\end{equation}
where we have employed the logarithmic derivatives $\Theta_{x_{i}} = x_{i} \partial_{x_{i}}$.

We will also make use of the rescaled coordinates
\begin{equation}
    \overline{x} = 2^{4} 3^{3} x,\quad \overline{y} = 2^{2} y,\quad \overline{z} = 2^{2} z.
\end{equation}

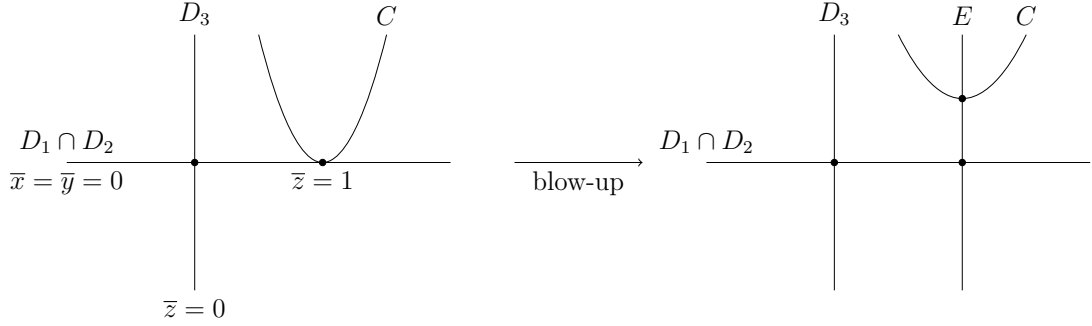
\begin{figure}[!htb]
    \centering
    \begin{adjustbox}{width=\textwidth}
    \begin{tikzpicture}
    \draw (0,0) node[above,midway] {$D_{1} \cap D_{2}$} node[below,midway] {$\overline{x} = \overline{y} = 0$} -- (6,0);
    \draw (2,2) node[above] {$D_{3}$} -- (2,-2) node[below] {$\overline{z} = 0$};
    \draw (3,2) parabola bend (4,0) (5,2) node[above] {$C$};
    \filldraw (2,0) circle (0.05);
    \filldraw (4,0) circle (0.05) node[below] {$\overline{z} = 1$};
    \draw[->] (7,0) -- node[below] {blow-up} (9,0);
    \draw (10,0) node[above] {$D_{1} \cap D_{2}$} -- (16,0);
    \draw (12,2) node[above] {$D_{3}$} -- (12,-2);
    \draw (14,2) node[above] {$E$} -- (14,-2);
    \draw (13,2) parabola bend (14,1) (15,2) node[above] {$C$};
    \filldraw (12,0) circle (0.05);
    \filldraw (14,0) circle (0.05);
    \filldraw (14,1) circle (0.05);
    \end{tikzpicture}
    \end{adjustbox}
    \caption{A simplified depiction of the moduli space of $\mathbb{P}_{1,1,2,8,12}[24]$, only looking at the divisors, intersections and tangencies of interest. The LCS point and the $(\overline{x},\overline{y},\overline{z}) = (0,0,1)$ conifold are shown.}
    \label{fig:P112812_blow-up}
\end{figure}

The two points of interest in the following are the LCS point $(\overline{x},\overline{y},\overline{z}) = (0,0,0)$ and the conifold point $(\overline{x},\overline{y},\overline{z}) = (0,0,1)$. We blow-up the conifold point by introducing exceptional divisors as schematically represented in figure \ref{fig:P112812_blow-up}, and we define coordinates
\begin{equation}
    x_{1} = \overline{x},\quad x_{2} = \frac{\overline{y}\overline{z}^{2}}{(1-\overline{z})^{2}},\quad x_{3} = 1 - \overline{z},
\label{eq:LCS_side_blow_up}
\end{equation}
at the LCS side of the blow-up and
\begin{equation}
    x_{1} = \overline{x},\quad x_{2} = 1 - \frac{\overline{y}\overline{z}^{2}}{(1-\overline{z})^{2}},\quad x_{3} = 1 - \overline{z},
\end{equation}
at the conifold side of the blow-up. We will work on the LCS side of the blow-up, but the final results are independent of the one chosen.

\subsubsection{An integral symplectic basis at the LCS}

We would like to obtain an integer symplectic basis at the LCS from $\omega_{\text{LCS}}$, the local basis of periods obtained from  \eqref{hyperbasis} and printed explicitly in appendix \ref{subsec:local_LCS_periods}. In practice, we need to calculate a transition matrix $m_{1}$ such that $\Pi = m_{1} \cdot \omega_{\text{LCS}}$ is an integer symplectic basis.

To this end we start by writing the prepotential at the LCS for $\mathbb{P}_{1,1,2,8,12}[24]$. The general expression for such a prepotential is \cite{Hosono:1994ax}
\begin{equation}
    \mathcal{F} = \frac{1}{6}K_{ijk}^{0}t^{i}t^{j}t^{k} + \frac{1}{2}a_{i,j}t^{i}t^{j} + b_{i}t^{i} + \frac{1}{2}c + \mathcal{F}_{\text{inst}}\,,
\label{eq:general_LCS_prepotential}
\end{equation}
where $c = -\frac{\zeta(3)}{(2\pi i)^{3}}\chi$ with $\chi$ the Euler
number of the manifold. The classical triple intersection numbers
$K_{ijk}^{0}$ are given in \cite{Hosono:1993qy}.
The $b_{i}$ are related to the intersections of the
second Chern class and the K\"ahler forms. Both the $b_{i}$ and $\chi$
can be calculated from the Mori-cone basis and the classical triple
intersection numbers through explicit expressions given in
\cite{Hosono:1994ax}. The $a_{i}$ are fixed modulo an irrelevant
integer part by demanding that the prepotential gives periods with
integer 
monodromies.

The resulting prepotential at the LCS for $\mathbb{P}_{1,1,2,8,12}[24]$ is
\begin{equation}
    \mathcal{F} = -\frac{4}{3}t_1^3-t_1^2 t_2-2 t_1^2 t_3-t_1 t_3^2-t_1 t_2 t_3+\frac{23}{6}t_1+t_2+2 t_3+240\frac{\zeta(3)}{(2\pi i)^{3}} + \mathcal{F}_{\text{inst}}\;.
\label{eq:LCS_prepotential}
\end{equation}
From it we obtain an integer symplectic basis of periods
\begin{equation}
    \Pi = ( 1, t^{1}, t^{2}, t^{3}, \partial_{t^{3}}\mathcal{F}, \partial_{t^{2}}\mathcal{F}, \partial_{t^{1}}\mathcal{F}, 2\mathcal{F} - t^{i}\partial_{t^{i}}\mathcal{F})\,.
\end{equation}
To calculate $m_{1}$ we match the leading behavior of $\Pi$ and
$\omega_{\text{LCS}}$. To this purpose we insert the leading terms of the
mirror maps  into $\Pi$
\begin{equation}
    t^{i} \sim \frac{1}{2\pi i}\log{x_{i}}\,,
\end{equation}
and work with a general ansatz for $m_{1}$. The latter  is constrained by demanding that the monodromies $M_{x_{i}}$ around the LCS divisors are compatible in both bases, i.e. $M_{x_{i}}^{\Pi} \cdot m_{1} = m_{1} \cdot M_{x_{i}}^{\omega_{\text{LCS}}}$. The resulting matrix is
\begin{equation}
\label{m1}
    m_{1}=
    \begin{pmatrix}
    1 & 0 & 0 & 0 & 0 & 0 & 0 & 0 \\
    0 & 1 & 0 & 0 & 0 & 0 & 0 & 0 \\
    0 & 0 & 1 & 0 & 0 & 0 & 0 & 0 \\
    0 & 0 & 0 & 1 & 0 & 0 & 0 & 0 \\
    2 & 0 & 0 & 0 & -2 & -1 & 0 & 0 \\
    1 & 0 & 0 & 0 & -1 & 0 & 0 & 0 \\
    \frac{23}{6} & 0 & 0 & 0 & -4 & -2 & -1 & 0 \\[0.1cm]
    \frac{60 i \zeta (3)}{\pi ^3} & \frac{23}{6} & 1 & 2 & 0 & 0 & 0 & 1 \\
    \end{pmatrix}\,.
\end{equation}

\subsubsection{Symplectic form}

To calculate the symplectic form (\ref{eq:symplectic_form}) for $\mathbb{P}_{1,1,2,8,12}[24]$ we start by writing the most general ansatz taking into account the order of the PF operators, which reads
\begingroup
\allowdisplaybreaks
\begin{align}
    Q = &A_{1}(\overline{x},\overline{y},\overline{z})\, 1 \wedge \Theta_{\overline{x}}
    + A_{2}(\overline{x},\overline{y},\overline{z})\, 1 \wedge \Theta_{\overline{y}}
    + A_{3}(\overline{x},\overline{y},\overline{z})\, 1 \wedge \Theta_{\overline{z}}\nonumber\\
    &+ A_{4}(\overline{x},\overline{y},\overline{z})\, 1 \wedge \Theta_{\overline{x}}\Theta_{\overline{y}}
    + A_{5}(\overline{x},\overline{y},\overline{z})\, 1 \wedge \Theta_{\overline{x}}\Theta_{\overline{z}}
    + A_{6}(\overline{x},\overline{y},\overline{z})\, 1 \wedge \Theta_{\overline{y}}\Theta_{\overline{z}}\nonumber\\
    &+ A_{7}(\overline{x},\overline{y},\overline{z})\, \Theta_{\overline{x}} \wedge \Theta_{\overline{y}}
    + A_{8}(\overline{x},\overline{y},\overline{z})\, \Theta_{\overline{x}} \wedge \Theta_{\overline{z}}
    + A_{9}(\overline{x},\overline{y},\overline{z})\, \Theta_{\overline{y}} \wedge \Theta_{\overline{z}}\nonumber\\
    &+ A_{10}(\overline{x},\overline{y},\overline{z})\, \Theta_{\overline{x}} \wedge \Theta_{\overline{x}}\Theta_{\overline{y}}
    + A_{11}(\overline{x},\overline{y},\overline{z})\, \Theta_{\overline{x}} \wedge \Theta_{\overline{x}}\Theta_{\overline{z}}
    + A_{12}(\overline{x},\overline{y},\overline{z})\, \Theta_{\overline{x}} \wedge \Theta_{\overline{y}}\Theta_{\overline{z}}\nonumber\\
    &+ A_{13}(\overline{x},\overline{y},\overline{z})\, \Theta_{\overline{y}} \wedge \Theta_{\overline{x}}\Theta_{\overline{y}}
    + A_{14}(\overline{x},\overline{y},\overline{z})\, \Theta_{\overline{y}} \wedge \Theta_{\overline{x}}\Theta_{\overline{z}}
    + A_{15}(\overline{x},\overline{y},\overline{z})\, \Theta_{\overline{y}} \wedge \Theta_{\overline{y}}\Theta_{\overline{z}}\nonumber\\
    &+ A_{16}(\overline{x},\overline{y},\overline{z})\, \Theta_{\overline{z}} \wedge \Theta_{\overline{x}}\Theta_{\overline{y}}
    + A_{17}(\overline{x},\overline{y},\overline{z})\, \Theta_{\overline{z}} \wedge \Theta_{\overline{x}}\Theta_{\overline{z}}
    + A_{18}(\overline{x},\overline{y},\overline{z})\, \Theta_{\overline{z}} \wedge \Theta_{\overline{y}}\Theta_{\overline{z}}\nonumber\\
    &+ A_{19}(\overline{x},\overline{y},\overline{z})\, 1 \wedge \Theta_{\overline{x}}\Theta_{\overline{y}}\Theta_{\overline{z}}\,.
\end{align}
\endgroup
Imposing then the constraints \eqref{eq:constancy_symplectic_form} that the symplectic form is constant throughout the moduli space when evaluated on solutions of the PF system we arrive at
expressions for the coefficients $A_i$. Their exact form is listed in appendix \ref{subsec:symplectic_form_coefficients}.

Inserting the integer symplectic basis of periods at the LCS that we obtained above into the symplectic form yields the intersection matrix
\begin{equation}
Q = 
    \begin{pmatrix}
    0 & 0 & 0 & 0 & 0 & 0 & 0 & \frac{i \eta }{16 \pi ^3} \\
    0 & 0 & 0 & 0 & 0 & 0 & \frac{i \eta }{16 \pi ^3} & 0 \\
    0 & 0 & 0 & 0 & 0 & \frac{i \eta }{16 \pi ^3} & 0 & 0 \\
    0 & 0 & 0 & 0 & \frac{i \eta }{16 \pi ^3} & 0 & 0 & 0 \\
    0 & 0 & 0 & -\frac{i \eta }{16 \pi ^3} & 0 & 0 & 0 & 0 \\
    0 & 0 & -\frac{i \eta }{16 \pi ^3} & 0 & 0 & 0 & 0 & 0 \\
    0 & -\frac{i \eta }{16 \pi ^3} & 0 & 0 & 0 & 0 & 0 & 0 \\
    -\frac{i \eta }{16 \pi ^3} & 0 & 0 & 0 & 0 & 0 & 0 & 0 \\
    \end{pmatrix}\,.
\end{equation}
This allows us to fix $\eta = -16 i\pi^{3}$ so that the symplectic form returns the correctly normalized intersections.

\subsubsection{A (numerical) integral symplectic basis at the conifold}

As described in section \ref{subsec:computing_local_periods} we can
generate a local basis of periods at the
$(\overline{x},\overline{y},\overline{z})=(0,0,1)$ conifold by
expressing the PF system in the \eqref{eq:LCS_side_blow_up}
coordinates and finding solutions to it order by order. A set of
solutions $\omega_{c}$ is given in appendix \ref{subsec:local_coni_periods}.

To transform $\omega_{c}$ into an integer symplectic basis at the
conifold we need to determine a suitable transition matrix as
explained in section \ref{subsec:integer_symplectic_basis}. Instead of
working with a fully general ansatz for this matrix, we already
restrict to the candidates among which the combinations giving the
$\alpha$-cycles will be chosen such that they do not mix with the would-be $\beta$-cycles, i.e. $(m_{2})_{1-4,5-8} = 0$.

Inserting $\Pi = m_{2} \cdot \omega_{c}$ into the symplectic form and demanding that the intersections are
\begin{equation}
    Q = 
    \begin{pmatrix}
    0 & 0 & 0 & 0 & 0 & 0 & 0 & 1\\
    0 & 0 & 0 & 0 & 0 & 0 & 1 & 0\\
    0 & 0 & 0 & 0 & 0 & 1 & 0 & 0\\
    0 & 0 & 0 & 0 & 1 & 0 & 0 & 0\\
    0 & 0 & 0 & -1 & 0 & 0 & 0 & 0\\
    0 & 0 & -1 & 0 & 0 & 0 & 0 & 0\\
    0 & -1 & 0 & 0 & 0 & 0 & 0 & 0\\
    -1 & 0 & 0 & 0 & 0 & 0 & 0 & 0\\
    \end{pmatrix}
\end{equation}
we find the relations that have to hold between the entries of $m_{2}$.

To proceed with the numerical matching we need to select points in the overlap of the regions of convergence of $\omega_{\text{LCS}}$ and $\omega_{\text{c}}$. Given the conditions \eqref{eq:constancy_symplectic_form} imposed on the symplectic form, the intersection of two periods given by it is a constant. Taking mixed intersections between the periods in $\omega_{\text{LCS}}$ and in $\omega_{\text{c}}$ we obtain functions that plateau in the region where the series expansions still correctly capture the behavior of both periods, thereby guiding us in the choice of points.

In this way we obtain that the transition matrix transforming $\omega_{c}$ into an integer symplectic basis is
\begin{equation}
    m_{2} =
    \resizebox{0.8\hsize}{!}{
    $\begin{pmatrix}
    1.00 & 0 & 0 & 0 & 0 & 0 & 0 & 0 \\
    1.08 i & -0.159 i & 0 & -0.159 i & 0 & 0 & 0 & 0 \\
    0.238 i & 0 & -0.318 i & -0.0194 i & 0 & 0 & 0 & 0 \\
    -0.00776 i & 0 & 0 & 0.327 i & 0 & 0 & 0 & 0 \\
    4.45 & -0.724 & -0.350 & -0.0527 & 0.0246 & 0.00150 & 0.0492 & 0 \\
    2.14 & -0.343 & 0 & 0.0179 & 0 & 0.0253 & 0 & 0 \\
    8.93 & -1.44 & -0.685 & 0.0103 & 0 & 0 & 0.101 & 0 \\
    4.80 i & 0.204 i & 0.0523 i & -0.0196 i & 0.000191 i & -0.00601 i & -0.109 i & 0.00538 i \\
    \end{pmatrix}$}
    \,.
\end{equation}
Both $\omega_{\text{LCS}}$ and $\omega_{\text{c}}$ were expanded to $\mathcal{O}(x^{11})$ to perform the matching, but in spite of this the convergence of the numerical values is still not enough, as some of the entries present an error of a few $\%$ when compared to the exact result \eqref{eq:analytic_m2} that we calculate below.

On top of this the result is very sensitive to slight changes in the choice of points. Shifting one of the points from $(\overline{x}, \overline{y}, \overline{z})$ to $(\overline{x}, \overline{y}-10^{-5}, \overline{z})$ the result changes noticeably, now being
\begin{equation}
    m_{2} =
    \resizebox{0.8\hsize}{!}{
    $\begin{pmatrix}
    1.00 & 0 & 0 & 0 & 0 & 0 & 0 & 0 \\
    1.07 i & -0.159 i & 0 & -0.158 i & 0 & 0 & 0 & 0 \\
    0.220 i & 0 & -0.318 i & 0 & 0 & 0 & 0 & 0 \\
    0.00292 i & 0 & 0 & 0.315 i & 0 & 0 & 0 & 0 \\
    4.49 & -0.720 & -0.343 & -0.0802 & 0.0256 & 0 & 0.0507 & 0 \\
    2.16 & -0.343 & 0 & 0.000337 & 0 & 0.0253 & 0 & 0 \\
    8.94 & -1.44 & -0.685 & -0.000187 & 0 & 0 & 0.101 & 0 \\
    4.77 i & 0.203 i & 0.0505 i & 0.00721 i & -0.0000747 i & -0.00558 i & -0.109 i & 0.00538 i \\
    \end{pmatrix}$}\,.
\end{equation}
These problems can be solved by an analytic determination of the transition matrix $m_{2}$.

\subsection{Analytic transition matrix}
In this section we will provide an analytic solution for the transition
matrix to the conifold  in the $\mathbb{P}_{1,1,2,8,12}[24]$ CY, which
is an example of a $\mathbb{P}_1$-fibration. This leads to expressions
for the periods in terms of hypergeometric $_3F_2$ functions and
derivatives thereof, which can be evaluated analytically, allowing us
to give an exact expression for the prepotential at the conifold, not
involving any factors which can only be determined numerically. This also
shows that all factors in the prepotential are rational numbers, a
fact important for our algorithm described in the next 
section.

\begin{figure}[!htb]
    \centering
    \begin{adjustbox}{width=\textwidth}
    \begin{tikzpicture}
    \node at (0,0) {$\omega$};
    \node at (5,0) {$\omega$};
    \node at (10,0) {$\omega_c$};
    \node at (0,5) {$\Pi$};
    \node at (10,5) {$\Pi$};
    \node at (0,-1) {$x_3=1$};
    \node at (10,-1) {$x_3=0$};
    \draw[<->] (1,0) -- node[below] {analytic continuation} (4.5,0);
    \draw[<->] (5.5,0) -- node[below] {coefficient matching}
    node[above] {m}(9.5,0);
    \draw[<->] (0,0.5) -- node[left] {monodromy} node[right] {$m_1$} (0,4.5);
     \draw[<->] (10,0.5) -- node[right] {$m_2=m\cdot m_3$} (10,4.5);
    \end{tikzpicture}
    \end{adjustbox}
    \caption{The different bases involved in the computation and the relations in between them.}
    \label{fig:maps}
\end{figure}
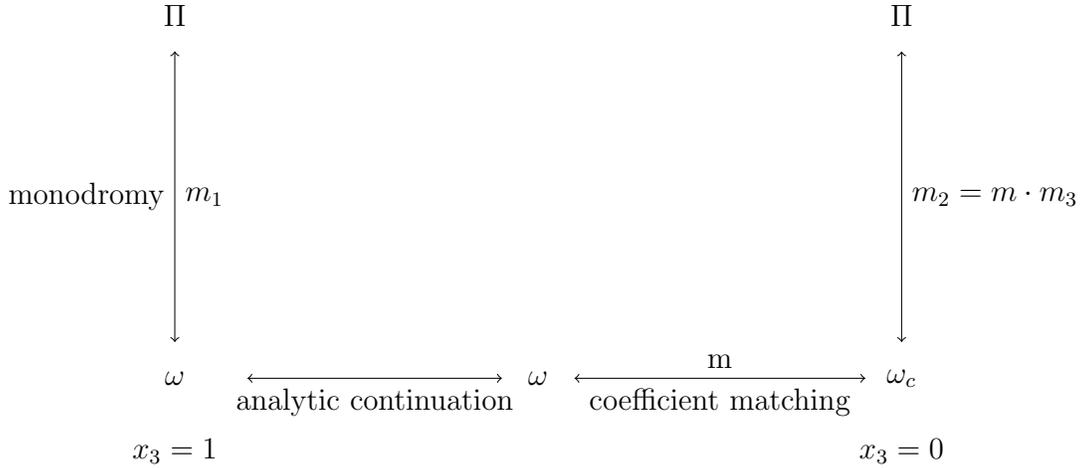

\newpage
To determine the prepotential to high orders we introduce several bases:
\begin{itemize}
	\item The symplectic basis $\Pi$.
	\item The hypergeometric basis $\omega$.
	\item The local PF basis around the conifold $\omega_c$.
\end{itemize}
The PF basis $\omega_c$ has the advantage that it is easy to evaluate to high order as described in the previous section. The hypergeometric basis can be related to the symplectic basis exactly. Moreover, it can be expanded around the conifold in terms of derivatives of hypergeometric functions, which allows us to match it exactly to the local basis. Combining these two transformations gives the relation between the local basis and the symplectic basis.
\begin{equation}
	\Pi=m_1 \cdot \omega=m_2 \cdot \omega_{c}\;.
\end{equation}
The relations between the different bases are shown in figure \ref{fig:maps}.
The hypergeometric basis $\omega$ is the local basis around the LCS \eqref{hyperbasis}. The transition matrix between this basis and the symplectic basis, $m_1$, can be determined purely on monodromy considerations around the LCS with the result \eqref{m1}.
For the analytic continuation to the conifold we rewrite the fundamental period in terms of a hypergeometric function. One can perform this sum for any coordinate, without loss of generality we choose the $z$ direction. We denote the l-vector corresponding to this direction $l^{(z)}$. The fundamental period then takes the form
\begin{equation}
\label{fundamental}
\omega_0=\sum_{n_1=0}^\infty\sum_{n_2=0}^\infty \ov{x}^{n_1+\rho_1}\ov{y}^{n_2+\rho_2}\ov{z}^{\rho_z}f(n_1,n_2,\rho_1,\rho_2,\rho_3)\;{}_pF_q({\vec{a}},{\vec{b}},\ov{z})\;,
\end{equation}
where $f$ denotes a complicated combination of $\Gamma$ functions independent of the coordinates and $\vec{a},\vec{b}$ are parameter vectors of length $p$ and $q$, depending on the $l$-vectors. Here
\eq{p=1+\sum\limits_{k, l^{(z)}_k<0}|l^{(z)}_k|\,,\quad\,q=\sum\limits_{k, l^{(z)}_k>0}|l^{(z)}_k|\,,}
i.e. $p$ is the sum  of the \emph{negative} entries of the charge row vector $l^{(z)}$ (plus 1), and $q$ is the sum of the \emph{positive} entries of the same row. 
Due to the CY-condition these sums have to be equal and $p=q+1$. The entries of $l^{(z)}$ appear inversely in the parameters $\vec{a}$ of the hypergeometric function. 
The exact form of the hypergeometric function is model dependent. 
The need to compute the derivatives with respect to the parameters later on imposes the computational constraint that no entry in $l^{(z)}$ may have an absolute value larger than 2
\begin{equation}
	|l^{(z)}_k|\le 2,\qquad \forall k\;,
\end{equation}
as otherwise rational parameters beyond $1/2$ appear. We now specialize to the charge row 
\begin{equation}
l^{(z)}=\left(0,\dots,0,1,1,-2\right)\;,
\end{equation}
where the ordering and number of zeros do not matter. This structure appears quite commonly, e.g. in the hypersurfaces $\mathbb{P}_{1,1,2,2,2}[8]$, $\mathbb{P}_{1,1,2,2,6}[12]$ and $\mathbb{P}_{1,1,2,8,12}[24]$ as well as in the complete intersections
$X_{(2|4)}(11|1111)$, $X_{(2|2|3)}(11|11|111)$ and $X_{(2|2|2|2)}(11|11|11|11)$ \cite{Hosono:1993qy}.
 The hypergeometric functions appearing in these manifolds when summing over the $z$ coordinate have only integer and half-integer parameters. To prevent clustering of the formulas we now specialize again to our example $\mathbb{P}_{1,1,2,8,12}[24]$, but the computation is similar for all such models. In this case the hypergeometric function in \eqref{fundamental} takes the form
  \begin{equation}
  \resizebox{0.88\hsize}{!}{$
 _3F_2\left(\left\{1,\rho_z-\frac{1}{2}\rho_1-\frac{1}{2}n_1 ,\frac{1}{2}+\rho_z-\frac{1}{2}\rho_1-\frac{1}{2}n_1 \right\},\left\{1+\rho_z,1+\rho_z-2\rho_2-2n_2\right\},\ov{z}\right)\;.
 $}
 \end{equation}
 The periods are now given by up to third order derivatives of this hypergeometric function with respect to its parameters. These can be evaluated e.g. using the HypExp2 package \cite{Huber:2007dx}. It has been proven in general that it is always possible to rewrite the generalized hypergeometric functions in terms of multiple polylogarithms \cite{Kalmykov:2008gq}, allowing us to express the derivatives in terms of harmonic polylogarithms (HPL). These can then be expanded\footnote{There is a technical subtlety involved in this expansion. The hypergeometric basis is actually divergent at the conifold. These divergences cancel out in the symplectic basis. Thus one has to first apply the transformation matrix $m_1$ before expanding.} around the conifold coordinates $x_i$ to any necessary order. The expansion in terms of  $x_1$ and $x_2$ coordinates has to be calculated term by term. While these can be in principle calculated to arbitrary order, the time needed to evaluate the derivatives becomes impractical already at low orders. Thus, we found it easier to calculate the transition matrix $m_2$ to the local basis and compute the instanton corrections in these coordinates.
Transforming the PF operators into these coordinates and using the Ansatz \eqref{eq:ansatz} gives the local periods. The matching of the coefficients in the expansion around the conifold uniquely fixes the transition matrix\\
\begin{equation}
    m_{2} = 
    \resizebox{0.80\hsize}{!}{
    $\begin{pmatrix}
	1 & 0 & 0 & 0 & 0 & 0 & 0 & 0 \\[0.1cm]
	\frac{i d}{2 \pi } & -\frac{i}{2 \pi } & 0 & -\frac{i}{2 \pi } & 0 & 0 & 0 & 0 \\[0.1cm]
	\frac{i \log (2)}{\pi } & 0 & -\frac{i}{\pi } & 0 & 0 & 0 & 0 & 0 \\[0.1cm]
	0 & 0 & 0 & \frac{i}{\pi } & 0 & 0 & 0 & 0 \\[0.1cm]
	\frac{a_7}{2} & \frac{-11 \log (2)-6 \log (3)}{2 \pi ^2} & -\frac{d}{2 \pi ^2} & \frac{1-3 \log (2)}{2 \pi ^2} & \frac{1}{4 \pi ^2} & 0 & \frac{1}{2 \pi ^2} & 0 \\[0.1cm]
	a_6 & -\frac{d}{2 \pi ^2} & 0 & 0 & 0 & \frac{1}{4 \pi ^2} & 0 & 0 \\[0.1cm]
	a_7 & \frac{-11 \log (2)-6 \log (3)}{\pi ^2} & -\frac{d}{\pi ^2} & 0 & 0 & 0 & \frac{1}{\pi ^2} & 0 \\[0.1cm]
	a_8 & b &c & 0 & 0 & -\frac{i \log (2)}{4 \pi ^3} & -\frac{i d}{2 \pi ^3} & \frac{i}{6 \pi ^3} \\
    \end{pmatrix}$}
    \,,
\label{eq:analytic_m2}
\end{equation}
 where 
 \begin{align*}
 a_6&=\frac{4 \pi ^2+25 \log ^2(2)+9 \log ^2(3)+30 \log (2) \log (3)}{4 \pi ^2}\;,\\[0.2cm]
 a_7&=\frac{23 \pi ^2+180 \log ^2(2)+54 \log ^2(3)+198 \log (2) \log (3)}{6 \pi ^2}\;,\\[0.2cm]
 a_8&=\frac{i \left(726 \zeta (3)-325 \log ^3(2)-54 \log ^3(3)-540 \log ^2(2) \log (3)\right)}{12 \pi ^3}\;,\\[0.2cm]
 &+\frac{\left(-297 \log (2) \log ^2(3)+127 \pi ^2 \log (2)+69 \pi ^2 \log (3)\right)}{12 \pi ^3}\;,\\[0.2cm]
 b&=\frac{i \left(-23 \pi ^2+180 \log ^2(2)+54 \log ^2(3)+198 \log (2) \log (3)\right)}{12 \pi
 	^3}\;,\\[0.2cm]
 c&= \frac{i \left(-4 \pi ^2+25 \log ^2(2)+9 \log ^2(3)+30 \log (2) \log (3)\right)}{4 \pi ^3}\;,\\[0.2cm]
 d&=5\log (2)+ 3\log (3)\;.
 \end{align*}
 While these expressions are rather long and non-rational, the important part is that all entries are known analytically, such that the cancellation of the irrational factors in the following steps is manifest.
 
 Applying this matrix to the local solution around the conifold gives an expression for the periods in the symplectic basis to arbitrary order. The periods themselves, especially those corresponding to the $\beta$-cycles, are too long to be presented here. After dividing by the fundamental period, the $\alpha$-periods which represent the mirror map take the form\\
\begin{equation*}
\resizebox{1\hsize}{!}{$
\begin{pmatrix}
 1 \\[0.1cm]
 \frac{\log \left(x_1\right)}{2 \pi i}-\frac{31 i x_1}{72 \pi }+\frac{i x_3}{4 \pi }+\frac{5 i x_1 \sqrt{x_3}}{72 \pi }-\frac{5 i x_1 x_2 \sqrt{x_3}}{1152 \pi }+\frac{i x_2 \sqrt{x_3}}{32 \pi }-\frac{i \sqrt{x_3}}{2 \pi }+\frac{3 i \log (3)}{2 \pi
   }+\frac{5 i \log (2)}{2 \pi } + \cdots \\[0.1cm]
 \frac{\log \left(x_2\right)}{2 \pi i}+\frac{x_3}{\pi i}-\frac{i \log \left(x_3\right)}{\pi }+\frac{i \log (2)}{\pi } + \cdots \\[0.1cm]
 \frac{5 i x_2 \sqrt{x_3} x_1}{576 \pi }-\frac{5 i \sqrt{x_3} x_1}{36 \pi }-\frac{i x_2 \sqrt{x_3}}{16 \pi }+\frac{i \sqrt{x_3}}{\pi} + \cdots
\end{pmatrix}.
$}
\end{equation*}
Changing the $x_3$ coordinate to $x_3=x_3^2$ and defining
 \eq{\label{qcoords}
     q_{U^1}&=864\, e^{2\pi i U^1}=x_1+\cdots\,,\\
     q_{U^2}&=\frac{4}{ (\frac{\pi}{i}Z)^4} \,e^{2\pi i U^2}=x_2+\cdots\,,\\
     q_Z\,&=({\textstyle\frac{\pi}{i}}Z)=x_3+\cdots
 }
allows us to invert the mirror map order by order. The numerical factors in the expressions for $q_{U^1}$ and $q_{U^2}$ arise from the chosen coordinates. If we had used the $x$, $y$ and $z$ coordinates instead of $\ov{x}$, $\ov{y}$ and $\ov{z}$ these would have been absent. The resulting mirror map is given by
 \begin{align}
     x_1&=q_{U^1}-q_{U^1} q_Z-\frac{31 q_{U^1}^2}{36}+\cdots\,,\\
     x_2&=q_{U^2}+\frac{5}{9}q_{U^1} q_{U^2} q_Z-\frac{q_{U^2}^2}{4}-\frac{5}{9} q_{U^1} q_{U^2}+\cdots\,,\\
     x_3&=q_Z+\frac{5}{36} q_{U^1} q_Z-\frac{5}{192} q_{U^1} q_{U^2} q_Z+\frac{1}{16} q_{U^2} q_Z+\cdots\,.
 \end{align}
 Moreover, the hypergeometric representation allows us to compute the mirror maps around the conifold exactly in the conifold coordinate. At $x_1=x_2=0$ the conifold modulus on the K\"ahler side is given by
\begin{align*}
	2\pi i Z&=\log(1-x_3^2)-2\log(2)+2\, \text{HPL}\left[-1;\frac{1-x_3}{1+x_3}\right]\\
	&=-2\,\textrm{arctanh}\!\left(x_3\right)\,.
\end{align*}
In this case the harmonic polylogarithm (HPL) actually reduces to a simple logarithm, but in higher orders more complicated HPLs of higher weight appear. In the appendix \ref{polylog} we give the basic definitions of harmonic polylogarithms. Finally, inserting the mirror map into the periods allows us to write down the prepotential around the conifold as 
\eq{
\label{prepotimp}
   \mathcal{F} = &-\frac{4}{3} (U^1)^3- U^2 (U^1)^2+\frac{23}{6} U^1+U^2  -\frac{120
     i}{\pi ^3} e^{2 i \pi  U^1}-\frac{35496 i}{\pi ^3} e^{4 \pi i  U^1}\\
   & -\frac{Z^3}{4}-2 (U^1)^2 Z -U^2 U^1 Z-U^1 Z^2+\frac{23}{12} Z +\frac{120}{\pi^2} e^{2 i \pi  U^1} Z\\
   & + Z^2\left(\frac{i  \log (2\pi Z )}{2 \pi }-\frac{3 i}{4 \pi }+\frac{1}{4}\right)+\frac{121 i \zeta (3)}{4 \pi ^3}+ \text{higher order}\,.
}
Note that all polynomial terms involving $U^1$ or $U^2$ are rational. The only non-rational terms are the quadratic $Z^2$ term and the constant $\zeta(3)$ term shown in the last row.
We also observe that the linear terms related to the $U^i$ are all given by the same topological numbers as they are in the LCS regime. The same holds for the manifold $\mathbb{P}^4_{1,1,2,2,6}[12]$. Together with the observation that the topological numbers at a conifold transition are given by sums of the LCS topological numbers \cite{Berglund:1995gd}, this would give rise to the conjecture, that all coefficients in the prepotential around the conifold except the quadratic terms are rational numbers. While we cannot prove this for the general case, it seams to be a rather frequent property.

Let us close this elaborate  mathematical section with a comment on possible generalizations.
If one wants to go beyond $\mathbb{P}^1$-fibrations, $\epsilon$-expansions for either hypergeometric $_3F_2$ with rational parameters or $_4F_3$ functions evaluated at $1$ are needed. These lead to much more complicated expressions in the transition matrices. For example in the 1-parameter complete intersection of four quadrics in $\mathbb{P}^7$ the fundamental period takes the form 
\begin{equation}
\omega_0(x)=\ _4F_3\left(\left\{\frac{1}{2},\frac{1}{2},\frac{1}{2},\frac{1}{2}\right\},\{1,1,1\},x\right)=\sum\limits_{n=1}^\infty \left[\frac{1}{4n}\dbinom{2n}{n}\right]^4x^n
\end{equation}
whose value at $1$ is given by \cite{rogers2013moments}
\begin{equation}
    \omega_0(1)=\frac{16}{\pi^2}L(f,2)\approx 1.118636\dots\;,
\end{equation}
the critical L-value of the weight four Hecke eigenform 
\begin{equation}
    f=\eta(2\tau)^4\eta(4\tau)^4\;,
\end{equation}
where $\eta(\tau)$ is the Dedekind $\eta$-function. This value will
appear in the entries of the transition matrix. 
In appendix \ref{lfunction} we provide  the basic definitions of
critical L-values. L-values are highly non-rational and for the given example even expressions in terms of $\Gamma$-functions are unknown\cite{rogers2013moments}. But many identities for ratios of L-values are known giving surprisingly simple, often rational, results. 
As an example, consider the weight 4 form
\begin{equation}
    f_2=\frac{\eta(4\tau)^{16}}{\eta(2\tau)^4\eta(8\tau)^4}\;,
\end{equation}
then the following identities hold \cite{rogers2013moments}:
\begin{equation}
  L(f_2,3)=\frac{\pi}{2}L(f,2)=\frac{\pi^2}{8}L(f_2,1)\;.
\end{equation}
Moreover, for critical values of a modular form $g$ it holds that 
\begin{equation}
    \frac{L(g,2k)}{L(g,2k-2)}=\text{algebraic number}\cdot \pi^s
\end{equation}
as well as
\begin{equation}
    \frac{L(g,2k+1)}{L(g,2k-1)}=\text{algebraic number}\cdot \pi^s
\end{equation}
for some integers k and s\cite{Shimura1977}. The algebraic numbers turn out to be rational in many cases, as e.g. in the example above.
Thus, our construction could also work in these cases, but this would require much more mathematical machinery which is beyond the scope of this paper.

\section{The quest for \texorpdfstring{$|W_0|\ll 1$}{|W0| << 1}}
\label{sec:quest_for_W0}

Now that we have developed the tools to calculate periods close to the conifold, we can continue towards the goal of this paper.
We want to investigate whether a method similar to that proposed by DKMM can be established in a region in moduli
space that is close to a conifold point.
After reviewing the construction at the LCS point as described by
DKMM, we will start off the conifold discussion by considering the
one-parameter model of  the quintic (or rather its mirror). 
Since this model has only one complex structure modulus, there is no direct
way of generalizing the DKMM construction, instead it turns out to be
rather a tuning problem whether fluxes can be chosen such that 
in the minimum $W_0\ll 1$. Indeed, one can find fluxes such that 
$W_0\approx 10^{-4}$, but to formulate a general mechanism geometries
with more complex
structure moduli are needed. As explicitly elaborated on  in the
previous section, in such  multi-parameter models there exits a regime, called Coni-LCS in the
following, which lies at the tangency between the conifold and the LCS
locus.
We will extend the DKMM construction to vacua close to the Coni-LCS regime 
and explicitly demonstrate the procedure using the $\mathbb{P}^{4}_{1,1,2,8,12}[24]$ example from section~\ref{CY-P112812}.

\subsection{Review of the DKMM construction}

Let us first briefly review the construction of Demirtas, Kim, McAllister and Moritz \cite{Demirtas:2019sip}.
The authors propose a two-step procedure to generate 
exponentially small $W_0$ terms at weak string coupling and 
large complex structure.
When using mirror variables, the prepotential splits into classical and non-perturbative terms.
Initially neglecting the non-perturbative terms, 
the first step is to find quantized fluxes 
for which the F-terms and superpotential vanish perturbatively.
DKMM formulate a Lemma which gives a sufficient condition to construct 
such solutions and directly determine the flat direction. 
In the second step, the previously neglected non-perturbative terms
generate a potential along the flat direction which can generically be stabilized 
to an exponentially small value by a racetrack-like procedure.

Let $X$ be an orientifold of a Calabi-Yau 3-fold with O3-planes and 
wrapped by D7-branes, carrying D3-brane charge $-Q_{\rm D3}$. 
With $\{A_a,B^b\}$ a symplectic basis of 3-cycles $H_3(X,\mathbb{Z})$: 
$A_a\cap A_b = B^a\cap B^b=0$,  $A_a\cap B^b=\delta_a^b$, 
the period vectors are defined as 
\eq{
	\Pi= \begin{pmatrix}  \int_{A_a}\Omega \\[0.1cm]\int_{B^a}\Omega \end{pmatrix}
	= \begin{pmatrix} X^a \\ \mathcal{F}^a  \end{pmatrix} .
}
$X^a$ are projective coordinates on the 
complex structure moduli space, and $\mathcal F$ is the prepotential with 
$\mathcal{F}_a=\partial_{X^a}\mathcal{F}$. We continue to work in a gauge where
$U^0=1$, so $\mathcal{F}_0=2\mathcal{F}-U^i\mathcal{F}_i$. 
From the 3-form field strengths $F_3$, $H_3$ one similarly  obtains the flux vectors 
\eq{
	F= \begin{pmatrix} \int_{A^a} F_3 \\[0.1cm] \int_{B_a} F_3 \end{pmatrix}\,,\quad
	H=\begin{pmatrix} \int_{A^a} H_3 \\[0.1cm] \int_{B_a} H_3	 \end{pmatrix} \,.
}
With the symplectic matrix 
$\Sigma=\begin{pmatrix} 0 & -1 \\ 1 & 0 \end{pmatrix}$ 
and $S$ the axio-dilaton,
the K\"ahler- and superpotential take the form
\eq{\label{DKMMpotentials}
	K&=	-\log\left(-i\,\Pi^\dagger\cdot\Sigma\cdot\Pi\right)
		-\log\left(S+\bar S\right)\,, 
	\\
	W&=	\left(F+iS H\right)^T\cdot\Sigma\cdot\Pi \,. 
}
When written in terms of the mirror variables, the tree-level
prepotential $\mathcal{F}$ can be separated into a classical, 
perturbative part $\mathcal{F}_{\rm pert}$ and 
non-perturbative instanton contributions\footnote{
It is important to notice that the ``non-perturbative" part in the 
mirror variables is part of the classical contribution to the type 
IIB theory. } $\mathcal{F}_{\rm inst}$, such that 
$\mathcal{F}(U)=\mathcal{F}_{\rm pert}(U)+\mathcal{F}_{\rm inst}(U)$ with
\eq{
	\mathcal{F}_{\rm pert}(U) &= 
		-\frac{1}{3!} K_{abc} U^a U^b U^c
		+\frac{1}{2} a_{ab} U^a U^b + b_a U^a + \xi \,,
	\\
	\mathcal{F}_{\rm inst}(U) &=
		\frac{1}{(2\pi i)^3}\sum_{\vec{q}}
			A_{\vec{q}}\; e^{2\pi i\, \vec{q}\cdot\vec{U}} \,.
}
The expressions refer to the mirror CY, so $K_{abc}$ are
the triple intersection numbers of the mirror, and the sum runs 
over effective curves in the mirror.
The constants $a_{ab}$, $b_a$ are rational numbers, and 
$\xi=-\frac{\zeta(3)\chi}{2(2\pi i)^3}$ with the Euler number 
$\chi$ of the CY.
The contributions to the superpotential stemming from 
$\mathcal{F}_{\rm pert}$ and $\mathcal{F}_{\rm inst}$ 
are respectively denoted $W_{\rm pert}$ and $W_{\rm inst}$, 
such that $W=W_{\rm pert} + W_{\rm inst}$.

Since the axionic real parts of $\vec{U}$ do not appear in the 
perturbative K\"ahler potential, they enjoy a 
discrete $\mathbb{Z}^n$ shift symmetry 
which is broken by generic fluxes. 
The shift symmetry generates a monodromy transformation on the flux vectors,
and only if such a monodromy combined with an appropriate
$SL(2,\mathbb{Z})$ transformation $(H,F)\to (H,F+rH)$, $r\in\mathbb{Z}$ 
leaves the flux vectors invariant there can be an 
unbroken remaining shift symmetry.

The first step then amounts to finding fluxes that allow for an 
unbroken shift symmetry, and finding moduli values that satisfy 
the F-flatness conditions with $W_{\rm pert}=0$. 
The following is a sufficient condition for the existence of such a 
\emph{perturbatively flat vacuum}.
If a pair of $\mathbb{Z}^n$ vectors $\vec{M}$, $\vec{K}$ exists such that
\begin{itemize}
	\item $-\frac{1}{2}\vec{M}\cdot\vec{K}\leq Q_{\rm D3}$,
	\item $N_{ab}=\mathcal{K}_{abc}\,M^c$ is invertible,
	\item $\vec{K}^T N^{-1} \vec{K}=0$,
	\item $\vec{p} = N^{-1} \vec{K}$ lies in the K\"ahler cone of the mirror CY,
	\item and $a\cdot\vec{M}$ and $\vec{b}\cdot\vec{M}$ are integer-valued,
\end{itemize}
then the fluxes
\eq{
\label{DKKMfluxes}
	F=\begin{pmatrix} \vec{b}\cdot\vec{M} \\ a\cdot\vec{M} \\ 0 \\ \vec{M} \end{pmatrix}
\qquad
	H=\begin{pmatrix} 0 \\ \vec{K} \\ 0 \\ 0 \end{pmatrix}
}
are compatible with the $Q_{\rm D3}$ tadpole bound, and the potential is 
perturbatively flat along $\vec{U}= \vec{p}\,S\,$ with $W_{\rm pert}|_{\vec{U}}=0$.

The non-perturbative contributions can now stabilize the remaining flat direction.
The effective superpotential along $\vec{U}$ in terms of the axio-dilaton $S$ is given at weak coupling by
\eq{
	\frac{W_{\rm eff}(S)}{\sqrt{2/\pi}} = M^a\partial_a\mathcal{F}_{\rm inst}
	=\sum_{\vec{q}} \frac{A_{\vec{q}} \vec{M}\cdot\vec{q}}{(2\pi i)^2}
	e^{2\pi i\vec{p}\cdot\vec{q} \,S}\,.
}
The final idea is to find flux quanta that stabilize $S$ via a race-track scenario,
balancing the two most relevant instantons $\vec{q}_1$, $\vec{q}_2$ against each other.
This is achieved when $\vec{p}\cdot\vec{q}_1 \approx
\vec{p}\cdot\vec{q}_2$. 

The conditions indicate that $h^{2,1}\geq 2$ is necessary in order to apply this mechanism. 
For a one-parameter model, the vectors and matrices are just numbers and $ K^2 N^{-1}=0$ means $K\!=\!0$. 
But then the perturbative vacuum found by the mechanism is $U= N^{-1}K\, S=0$ which is both 
outside the LCS regime of validity and has no flat direction 
along which the non-perturbative terms could generate a small $|W_0|$.

For a complete stabilization of all moduli, the hope is to continue with a KKLT-like procedure starting with this small $W_0$.
Unfortunately it is not quite so straightforward, as examples show that the perturbatively flat direction produces a mass scale of order $|W_0|$, which coincides with the mass scale of the K\"ahler moduli in the KKLT scenario.
The low energy theory must contain not only the K\"ahler moduli, but also the axio-dilaton, and the Pfaffian prefactors which appear in the non-perturbative superpotential cannot be treated as a constant.
DKMM argue that under some assumptions, the unbroken shift symmetry of the perturbatively flat vacuum would guarantee that the contributions of the axio-dilaton to the Pfaffian factors are exponentially small. Then one could reasonably approximate the Pfaffians by constants. To show this explicitly is however left open, and will also not be treated in our work.

\subsection{\texorpdfstring{$|W_0|\ll 1$}{|W0| << 1}  in the conifold regime}

For really getting the uplifted dS minimum in the last step of KKLT, 
one needs a strongly warped throat.
Thus, one needs a similar construction  in the
region close to a conifold point. This is not straightforward, 
as the periods take a completely
different form when expanded around such a point.

To set the stage, let us consider the simplest model,
namely the (mirror of the) quintic that has just a single complex
structure modulus. Close to the conifold point 
the period vector $\Pi^T=(X^0,X^1,F_0,F_1)$ can  be expressed as
\cite{Candelas:1990rm,Curio:2000sc,Huang:2006hq,Bizet:2016paj,Blumenhagen:2016bfp}
\eq{
\label{periods}
\Pi=X^0\,\left( \begin{matrix} 
1\\ Z\\ A +B Z+O(Z^2) \\ -  \frac{1}{2\pi i} Z\log Z + C + DZ + O(Z^2) 
\end{matrix}   \right)
}
with  parameters
\eq{
A&=(-0.103412 + 0.090045i)\,,\qquad
D=- (0.043170 - 0.039843i)\,,\\
B&=C = (0.074533 + 0.085597i)\,,
}  
that are only known numerically\footnote{There are known expressions
  for the transition matrix of all hypergeometric 1-parameter models
  in terms of L-values/quasiperiods of Hecke eigenforms of $\Gamma_0(N)$
  \cite{Klemm2}.}. Note that these are in general
irrational numbers though featuring certain correlations and
rationality properties. The relation $B=C$ is a consequence of the
existence of a prepotential for these periods, which reads
\eq{
         {\cal F}=-\frac{1}{4\pi i} Z^2 \log Z +\frac{A}{2} +BZ +\left(
           \frac{D}{2}+\frac{1}{8\pi i}\right)Z^2 +O(Z^3)\,.
}
The corresponding K\"ahler potential for the complex structure modulus
is given by
\eq{
\label{Kahlerconi}
           K_{\rm cs}&=-\log\left[ -i \Pi^\dagger \Sigma\, \Pi \right]\\[0.1cm]
    &=-\log\left[ \frac{1}{2\pi} |Z|^2\, \log( |Z|^2) + 2\Im(A) +
      2\Im(B)(Z+\ov Z) +\cdots\right].
}
This will be the leading order K\"ahler potential
in the volume-dominated regime, i.e. for ${\cal V} |Z|^2\gg 1$. Including also
the overall K\"ahler modulus $\cal V$ and the axio-dilaton $S$, the total
unwarped K\"ahler potential becomes
\eq{
\label{Kahlerconiunwarp}
            K_{\rm unwarp}=-2\log({\cal V}) - \log(S+\ov S) &-\log(2\Im(
            A)) -\frac{ \Im(B)}{\Im (A)} (Z+\ov Z)\\
                 &-\frac{1}{4\pi \Im(A)} |Z|^2\, \log( |Z|^2) +\cdots\,.
}
For the strongly warped, throat-dominated regime ${\cal V} |Z|^2\ll 1$, the effective
action was derived in
\cite{Douglas:2007tu,Bena:2018fqc,Blumenhagen:2019qcg}. 
Here the warping backreacts non-trivially so
that the K\"ahler potential  takes  the different form
\eq{
\label{Kahlerconiwarp}
   K_{\rm warp}&=-2\log({\cal V}) - \log(S+\ov S)+\xi \left( {|Z|\over {\cal V}}\right)^{2\over 3}\,,
}
with $\xi=c' M^2$, $c'$ an order one parameter and $M$ denoting
the  $F_3$ flux along the conifold A-cycle.
This K\"ahler potential features a warped no-scale structure
\eq{
\label{noscalea}
    \sum_{I,\ov J} G^{I\ov J} \partial_I K \partial_{\ov J} K
    = 3-(N-1)
    {\xi \vert Z\vert^{2\over
      3}\over {\cal V}^{2N\over 3}} + O(\xi^2) 
}
where the sum runs over the set $I,J\in\{T,Z\}$. Thus, precisely for
the K\"ahler potential \eqref{Kahlerconiwarp} the order $O(\xi)$ term vanishes.

\subsubsection{Moduli stabilization}

A general flux induced superpotential
\eq{
               W&= \int_{\mathcal M} \bigl( F +i S\, H \bigr) \wedge
               \Omega_3\,\\[0.1cm]
                      &=(X^\Lambda f_\Lambda - F_\Lambda \tilde f^\Lambda)
                     +iS(X^\Lambda h_\Lambda - F_\Lambda \tilde h^\Lambda)
}
leading to the stabilization of the conifold modulus at exponentially
small values can be expanded as
\eq{
              W&=-  {M\over 2\pi i} Z\log Z + \sum_{n=0}^\infty  M_n
              Z^n + iS  \sum_{n=0}^\infty  K_n Z^n \\
    &= -  {M\over 2\pi i} Z\log Z +M_0 +M_1 Z+iK_0 S  +iK_1 SZ+ {\cal O}(Z^2)\,,
}
with 
\eq{
     M&=-\tilde f^1\,,\quad M_0=f_0-A\tilde f^0-C \tilde f^1\,,\quad
     M_1=f_1-B \tilde f^0-D \tilde f^1\,,\\[0.1cm]
    K_0&=h_0 - A \tilde h^0 \,,\quad 
      K_1=h_1-B \tilde h^0\,.
}
Here we have chosen $\tilde h^1=0$ in order to avoid $(SZ \log Z)$-terms.
Note that while the quantized fluxes are integers, the coefficients $M_n$ and
$K_n$ are in general complex numbers.

Next we have to solve the minimum conditions $D_Z W=D_S W=0$. 
Using the K\"ahler potential \eqref{Kahlerconiunwarp}, one finds for
the volume-dominated case
\eq{
\label{partZW}
     D_Z W&= \partial_Z W +\partial_Z K\, W\\
     &= -{M\over 2\pi i} \log Z -{M\over 2\pi
       i}+M_1 +iK_1 S -{\Im(B)\over \Im(A)} \big(M_0+iK_0 S\big)+\cdots\,.
}
As shown in \cite{Blumenhagen:2019qcg},
in the warped, throat-dominated case,  the warped no-scale structure \eqref{noscalea}  implies
that the minimum of the scalar potential is at $\partial_Z W\approx
0$. This gives the same result as in \eqref{partZW} once we formally
set $\Im(B)=0$.

Solving \eqref{partZW}, in both cases at leading order the $Z$ modulus
can be written as
\eq{
           Z_0=\zeta_0 \exp\left(-{2\pi \hat K_1\over
               M}S_0\right)\,,\qquad 
         \zeta_0=\exp\left(2\pi i{\hat M_1\over M}\right)\,,
}
with parameters
\eq{
        \hat K_1=\begin{cases}  K_1-{\Im(B)\over \Im(A)} K_0 &
          \text{ volume-dominated} \\
             K_1 & \text{ throat-dominated} \end{cases}
}
and
\eq{
        \hat M_1=\begin{cases}  M_1-{M\over 2\pi i}-{\Im(B)\over \Im(A)} M_0 &
          \text{ volume-dominated} \\
              M_1-{M\over 2\pi i}  & \text{ throat-dominated}\,. \end{cases}
}
For $\hat K_1>M$ and $\Re(S)>1$ the value of the conifold modulus can be guaranteed to be
exponentially small, hence making our expansion in orders of $Z$
self-consistent.

Looking at the axio-dilaton condition $D_S W=0$,   at leading order we find
\eq{
\label{dscond}
         0=  iK_0 &+iK_1 Z \\
              &-{1\over S+\ov S}\left(  M_0 +iK_0 S +{M\over
               2\pi i} Z + {\Im(B)\over \Im(A)} \big(M_0+K_0 S\big) Z\right)\,,
}
where $D_ZW=0$ was invoked.
As in \cite{Blumenhagen:2016bfp}, for the stabilization of the axio-dilaton we now distinguish the two
cases, $K_0\ne 0$ and $K_0=0$.

\subsubsection*{Case A: $K_0\ne 0$}

In this case, the terms linear in $Z$ in \eqref{dscond} can be
neglected so that one gets the simple solution
\eq{
                  \ov S_0=-i{M_0\over K_0}\,.
}
For $\Re(S)\gg 1$ we need to require 
\eq{
\label{posdila}
             1\ll \Im(M_0/K_0)={M_0 \ov K_0 - \ov M_0 K_0\over 2i\,|K_0|^2} .
}
For the resulting value of the superpotential in the minimum one obtains
\eq{
\label{W0quintic}
            W_0=  \underbrace{M_0 \ov K_0 - \ov M_0 K_0\over \ov K_0}_{w_0} +O(Z_0)\,.
}
Thus, in order to have an exponentially small value of the
superpotential in the minimum, the leading order term $w_0$ in \eqref{W0quintic}
must vanish or at least  be very tiny. Thinking of $M_0$ and $K_0$ as
two-dimensional
vectors,  the superpotential $w_0$ vanishes if $M_0$ and $K_0$ are collinear.
Since $M_0$ and $K_0$  generically contain model dependent complex
valued parameters, 
solving this condition for the fluxes becomes a number theoretic question.

Let us analyze this in more detail using  the concrete values for the
(mirror of the) quintic. First one realizes that due to
\eqref{posdila}  $w_0=0$ implies $\Re(S)=0$
which means the
string coupling is infinitely large and thus  outside the
regime of validity. 
Moreover, using $w_0=2i\Re(S) K_0$ and $\Re(S)>1$ one can derive the lower bound
\eq{
                |w_0|>2 |K_0|=2 |h_0 - A \tilde h^0|>2 |\Im(A)|=0.18\,,
}
where we used that due to $K_0\ne 0$ not both $h_0$
and $\tilde h^0$  are allowed to vanish.
Thus, at least for the specific case of the quintic, in Case
A the superpotential in the minimum is bounded from below
by $|w_0|>O(10^{-1})$.  

\subsubsection*{Case B: $K_0=0$}

This means that we have $h_0=\tilde h^0=0$ so that $\hat K_1=K_1=h_1$ and
$M=-\tilde f^1$ are both integers.
Now, up to order $O(Z)$ the condition \eqref{dscond} reads
\eq{
\label{dscondb}
           iK_1 Z -{1\over S+\ov S}\left(  M_0  +{M\over
               2\pi i} Z\right)=0\,
}
where $Z$ is related to $S$ as $Z=\zeta_0 \exp(-{2\pi K_1\over M} S)$.
We observe that \eqref{dscondb} is nothing else than the vanishing
F-term condition $F_S=0$ for an effective superpotential 
\eq{
     W_{\rm eff}=M_0 +{M\over 2\pi i} \zeta_0\, e^{-{2\pi K_1\over M} S}\,.
}
This is very reminiscent
of the KKLT superpotential, where here we are dealing with a 
no-scale potential. 
Writing $S=s+ic$ one obtains for the $C_0$ axion
\eq{
\label{minic}
         c=-{M\over 2\pi K_1} \arg\left({M_0\over i\zeta_0}\right)
}
and the dilaton is given by the solution of the transcendental
equation
\eq{
\label{minis}
             \left\vert {M_0\over \zeta_0}\right\vert=\Big(2K_1
               s+{\textstyle {M\over
               2\pi}}\Big) e^{-{2\pi K_1\over M} s}\,.
}
As in KKLT this only admits a solution in the controllable regime if
the left hand side is very tiny, $M_0\ll 1$. Whether the flux landscape admits
such values is a model dependent  number theoretic question. Let us recall the parameters
\eq{
              M_0=f_0-A\tilde f^0-C \tilde f^1\,,\quad
              M_1=f_1-B \tilde f^0-D \tilde f^1\,,\quad M=-\tilde f^1
}
which are in general complex valued. One can easily convince oneself
that for the quintic there exist choices of the fluxes that yield $M_0=O(10^{-4})$, as
for instance
\eq{
                f_0=14\,,\qquad \tilde f^0=77   \,,\qquad \tilde
                f^1=-81\,.
}  
This gives $M_0\approx -(1+i) \cdot 10^{-4}$,  $M_1\approx -3.2 -3.4i$
and $M=81$. Moreover, one gets $\zeta_0\approx 0.46 -0.12i$.
For this choice the solution to \eqref{minic} and  \eqref{minis} is
\eq{
                  c_0\approx -{33.9\over K_1}\,,\qquad s_0\approx {180.4\over K_1}\,,
}
which for small enough $K_1$ is in a perturbative regime.
For the value of the conifold modulus we find $|Z_0|\sim 4\cdot
10^{-7}$ and
the value of the superpotential in the minimum is of the order of
$M_0$ namely
\eq{
                  |W_0|\sim |M_0|\approx 1.4\cdot 10^{-4}\,.
} 
Therefore, the Case B provides a controlled KKLT-like stabilization of the
complex structure and axio-dilaton moduli giving for the quintic a
Minkowski minimum  of the no-scale scalar potential
with a small value of $|W_0|$. This value was dialed by a suitable
choice of flux quantum numbers. In our case these were of the order $O(10^2)$
and so that there is   the concern of overshooting in some tadpole
cancellation conditions. In the example, there will a contribution to
the D3-brane tadpole  $Q_{\rm D3}=h_1 \tilde f^1=-K_1 M=O(10^2)$.

\subsubsection{Moduli masses}

The latter result is encouraging for extending the model \`a la KKLT by
adding a non-perturbative contribution to the superpotential
that depends on the K\"ahler modulus $T$. Recall that in the DKMM
construction the issue arises that the mass of the lightest complex structure
modulus is of the same order as the mass of the K\"ahler modulus,
calling for a more detailed analysis. Let us see how the situation is
in the conifold regime.

For estimating the masses, we compute the Hessian
$V_{ab}=\partial_a \partial_b V$ in the minimum, which for a no-scale model
simplifies considerably. Since $F_I=0$ in the minimum, the only 
non-vanishing contributions can come from 
\eq{
    \partial_a \partial_b V=e^K \left( K^{I\ov J} (\partial_a D_I W)
      (\partial_b D_{\ov J}
      \ov W) + (a\leftrightarrow b)\right)\,.
}
The masses in the canonically normalized field basis are the
eigenvalues of the matrix $K^{ac} V_{cb}$, where $K^{ac}$ denotes the
inverse  K\"ahler metric.

In the volume-dominated regime, we find for the mass eigenvalues the following
scaling\footnote{In the more precise relations also factors of the
  dilaton and the fluxes appear, but they do not change our conclusion.} with ${\cal V}$ and $|Z|$
\eq{
           m_Z^2\sim {M_{\rm pl}^2\over {\cal V}^2 |Z|^2}\sim {M_{\rm
               s}^2\over {\cal V} |Z|^2}\,,\qquad
        m_S^2\sim {M_{\rm pl}^2\over {\cal V}^2}\,.
}
In Case B we also have the relation $|Z|\sim M_0/s$. 
The expression for the mass  $m_Z$ makes it evident that the expressions in
this regime can only be valid for ${\cal V} |Z|^2\gg 1$, because otherwise
the mass of the conifold modulus would come out larger
than the string scale. Moreover, one always finds the hierarchy
$m_Z\gg m_S$. 
Extending this model to KKLT by also including a non-perturbative
contribution $A\exp(-a T)$  depending on the overall K\"ahler modulus, the mass
of the latter scales 
as
\eq{   
\label{masskaehlerm}
                             m_\tau^2\sim {|W_0|^2\over {\cal V}^{2\over
                                 3}}  M_{\rm pl}^2 \sim {|M_0|^2\over {\cal
                                 V}^{2\over 3}}  M_{\rm pl}^2\sim {|Z|^2\over {\cal
                                 V}^{2\over 3}}  M_{\rm pl}^2\,,
}
which for small $M_0$ can be kept much smaller than the complex
structure and axio-dilaton moduli.

Next consider the throat-dominated regime, where 
for Case A we find  the mass eigenvalues 
\eq{
           m_Z^2\sim   \left( {|Z|\over {\cal V}}
           \right)^{2\over 3} M_{\rm pl}^2\sim \big({\cal V} |Z|^2\big)^{1\over
             3} M_{\rm s}^2\,,\qquad
        m_S^2\sim {M_{\rm pl}^2\over {\cal V}^2} \,.
}
The expression for $m_Z$ nicely shows  that we need ${\cal V}
|Z|^2\ll 1$ in order for the mass to be smaller than the string scale.
Moreover, one has the hierarchy $m_S\gg  m_Z$. However, 
at least  for the concrete example of the quintic  we do not get
$|W_0|\ll 1$ in Case A.

For Case B there is an important change in the mass scales
\eq{
           m_Z^2\sim   \left( {|Z|\over {\cal V}}
           \right)^{2\over 3} M_{\rm pl}^2\,,\qquad
        m_S^2\sim \left( {|Z|\over {\cal V}}
           \right)^{4\over 3} M_{\rm pl}^2\,
}
so that now we have the inverted hierarchy $m_Z\gg m_S$. In
addition, taking into account  \eqref{masskaehlerm} for sufficiently small $|Z|$ the K\"ahler modulus can be kept
lighter than the axio-dilaton, i.e. $m_S\gg m_\tau$. 

This looks very promising, so let us summarize our findings:
In Case B, by a suitably tuned choice of fluxes one can stabilize the conifold modulus
 and the axio-dilaton in the controlled regime such that $|W_0|\sim
O(10^{-4})$ and their masses are hierarchically larger than the mass
of the K\"ahler modulus. Thus, the AdS KKLT minimum seems to
exist. In the throat-dominated regime, there is also a  tiny warp factor that
in principle could allow to  uplift the minimum to dS. However,  
in this case other issues might appear, like
the appearance of light KK modes localized at the tip of the long
throat, whose mass has been shown \cite{Blumenhagen:2019qcg} to scale like the mass of the $Z$ modulus.
This might spoil the validity of the employed effective action of just
the conifold modulus and the axio-dilaton.

While in the simple one-parameter model we could explore the stabilization of the conifold modulus, 
generalizing the DKMM procedure requires more moduli to work with. That Case B with $h_0=\tilde h^0=0$
showed more promise is nice, since these fluxes are also suggested by the procedure of DKMM.
In the following we shall propose a general algorithm which extends the work of DKMM to the Coni-LCS regime of a multi-parameter CY.

\subsection{\texorpdfstring{$|W_0|\ll 1$}{|W0| << 1} in the Coni-LCS regime}
\label{subsec:coniLCS-algorithm}

Consider an $n$-parameter CY with one modulus close to the conifold described in terms of the perturbative prepotential and instanton series
\eq{
\label{ansatzprepconi}
\mathcal{F}_{\rm pert}&=-\frac{1}{3!}
K_{ijk}X^iX^jX^k+\frac12A_{ij}X^iX^j+B_iX^i+C-\frac{Z^2\log Z}{2\pi i}\,,
\\
\mathcal{F}_{\rm inst}&= \frac{1}{(2\pi i)^3}\sum_{\vec c} a_{\vec c}\prod_{i=1}^n {q_i}^{n_i}\,,
}
with $q_i$ the coordinates used to invert the mirror map\footnote{
Since we are close to the conifold these coordinates are not simply exponentials of the moduli as in the LCS regime, 
but rather the conifold modulus enters linearly \eqref{qcoords}.} and $\vec c$ running over effective curves. 
To simplify  notation, we use latin indices to denote all moduli ${X^i=(\vec{U},Z)^T}, \, i=1,\ldots,n$, 
and greek indices to denote only the LCS   moduli $U^\alpha, \, \alpha=1,\ldots,n-1$.
If a pair of $\mathbb{Z}^{n}$ flux vectors $\vec{\tilde f}$, $\vec{h}$ exists such that
\begin{itemize}
    \item $-\frac12 \vec{\tilde f} \cdot \vec{h} \leq Q_{\rm D3}$,
	\item $N_{\alpha\beta}=K_{i\alpha\beta}\tilde f^i$ is invertible,
	\item $(N^{-1})^{\alpha\beta}h_\alpha h_\beta=0$,
	\item $p^\alpha = (N^{-1})^{\alpha\beta} h_\beta$ lies in the K\"ahler cone of the mirror CY,
	\item $A_{i\alpha}\tilde f^i$ and $B_i \tilde f^i$ are integer-valued,
\end{itemize}
then the fluxes
\eq{
\label{fluxes}
	F=\begin{pmatrix} B_i \tilde f^i \\ (A_{i\alpha}\tilde f^i,\,  f_n)^T\\ 0 \\ \vec{\tilde f}\end{pmatrix}\,,
\qquad
	H=\begin{pmatrix} 0 \\ \vec{h} \\ 0 \\ 0 \end{pmatrix}
}
are compatible with the $Q_{\rm D3}$ tadpole bound, and there is a \emph{perturbatively flat vacuum}  along 
\eq{ U^\alpha= p^\alpha\,S \,, \qquad Z=\zeta_0\,e^{-2\pi \frac{K_1}{M} S}
}
with $\zeta_0=e^{2\pi i\frac{M_1}{M}-1}$ and
\eq{
\label{theorempara}
M=-2\tilde f^n \,,\quad
M_1= f_n - A_{ni}\tilde f^i +\frac{\tilde f^n}{2 \pi i} \,,\quad
K_1=h_{n}- K_{i\alpha n}\tilde f^i (N^{-1})^{\alpha\beta} h_\beta 
}
along which $W_{\rm pert}|_{\vec{U},Z}\approx \frac{ Z M}{2\pi i}$ is exponentially small in $\Re(S)$.
As before, the conditions imply that too few moduli break the mechanism. Here , $h^{2,1}\geq3$ is necessary.
\\

Following a three-step procedure, let us outline in more detail how  this works. The periods are
computed from the prepotential as 
\eq{
X^0&=1,\quad X^\alpha=U^\alpha,\quad X^n=Z,\\[5pt]
F_0&= 2C+ B_iX^i +\frac{1}{3!}K_{ijk}X^iX^jX^k+\frac{ Z^2}{2\pi i}, 
\\[0.1cm]
F_i&= -\frac12 K_{ijk}X^jX^k +A_{ij}X^j+ B_i - \delta_{in}\left(\frac{Z}{2\pi i}+\frac{Z\log(Z)}{\pi i}\right)\,.
}
By restricting our choice of fluxes to 
\eq{
\tilde h^\Lambda=(0,0),
\quad h_\Lambda=(0,h_i),
\quad \tilde f^\Lambda=(0,\tilde f^i), 
\quad f_\Lambda = (B_i\tilde f^i, A_{\alpha i}\tilde f^i, f_n)\,
}
we obtain a superpotential which, similar to the DKMM case,  is homogeneous of order two at $Z=0$. 
Note that for this to work, $B_i\tilde f^i,\, A_{\alpha i}\tilde f^i$
must be {\it integer} valued, which calls for the parameters $A_{ij}$
and $B_i$ in the  prepotential \eqref{ansatzprepconi} to be {\it rational} numbers.
The resulting superpotential  can be expanded as
\eq{
W&=(F+iSH)^T\cdot\Sigma\cdot\Pi=(X^\Lambda f_\Lambda -F_\Lambda \tilde f^\Lambda)+ iS(X^\Lambda h_\Lambda -F_\Lambda \tilde h^\Lambda)\\[0.1cm]
&=\frac12 K_{ijk}\tilde f^iX^jX^k+\frac{\tilde f^n Z}{2\pi i}+\frac{\tilde f^n Z\log(Z)}{\pi i} +ih_iX^i S 
+(f_n-A_{ni}\tilde f^i)Z \,.
}
To proceed,  at zeroth order in $Z$ we first  stabilize the $U^\alpha$ moduli in a
supersymmetric minimum with  vanishing superpotential
\eq{
W=\frac12 N_{\alpha\beta}U^\alpha U^\beta +iSh_\alpha U^\alpha &=0\,, \\
\partial_\alpha W&=0\,,
}
with $N_{\alpha\beta}=K_{i\alpha\beta}\tilde f^i$.
Provided $N_{\alpha\beta}$ is invertible, the minimum is located at 
\eq{\label{Umin}
U^\alpha=p^\alpha S=-i S (N^{-1})^{\alpha\beta} h_\beta\,.
}
Demanding that $W=0$ results in a condition on the fluxes, $(N^{-1})^{\alpha\beta}h_\alpha h_\beta=0$.

Integrating out the moduli $U^\alpha$, since we invoked a vanishing superpotential at zeroth order in $Z$, 
the remaining terms of the  superpotential are at least of order $Z$ 
\eq{
W_{\rm pert}(S,Z)=-\frac{M Z\log(Z)}{2\pi i}+M_1 Z + iK_1SZ+O(Z^2)\,,
}
with the parameters given in \eqref{theorempara}.
For the F-term  we find
\eq{
D_ZW&=\partial_ZW+\partial_ZK \cdot O(Z) \\&= -\frac{M}{2\pi i}\log(Z)-\frac{M}{2\pi i}+M_1+iK_1S +O(Z)
}
showing that  the K\"ahler potential contribution to $D_ZW$ is of subleading order.
Thus,  the conifold modulus is stabilized at
\eq{\label{conifold-min}
Z_0=\zeta_0\,e^{-2\pi \frac{K_1}{M} S} \,,\quad \text{ with } \zeta_0=e^{2\pi i\frac{M_1}{M}-1}\,.
}
What we have found is a perturbatively flat vacuum extending the Lemma of DKMM, where the complex structure  moduli are stabilized in terms of the axio-dilaton as $\log(Z)\sim U^\alpha\sim S$. 

The final step is to integrate out $Z$, resulting in an effective superpotential  composed of the instanton superpotential $W_{\rm inst}=-\tilde f^i \partial_i \mathcal{F}_{\rm inst}$ 
as well as the linear  corrections in $Z$ resulting from  $W_Z=W_{\rm pert}|_{Z=Z_0}=\frac{ Z M}{2\pi i}$,
\eq{\label{Weff}
W_{\rm eff}=-\tilde f^i \partial_i \mathcal{F}_{\rm inst}+\frac{Z M}{2\pi i} \sim \sum a_n \, e^{c_n S} \,.
}
Similar to DKMM, such an effective non-perturbative superpotential has the potential
to stabilize the axio-dilaton by choosing fluxes that balance the
leading terms against each other  in a racetrack-like way. 
As long as the approximations we did along the way hold true in the minimum, 
the resulting $W_0$ can be stabilized at exponentially small values.
Here it is important to keep the instanton series under control, as 
the conditions $|q_i|<1$ will result in non-trivial constraints on the fluxes we may choose.

\subsection{Example: \texorpdfstring{$\mathbb{P}_{1,1,2,8,12}[24]$}{P(1,1,2,8,12)[24]}}

Now let us apply this generic algorithm to the example
$\mathbb{P}_{1,1,2,8,12}[24]$ worked out in detail in section  \ref{CY-P112812}.
Recall the form of the prepotential \eqref{prepotimp}, from which one
can read off the data for the perturbative part
\eq{
K_{111}&=8,\, K_{112}=2,\, K_{113}=4,\, K_{123}=1,\, K_{133}=2\,, \\[0.1cm]
A_{33}&=\left(\frac{1}{2} + \frac{3-2\log(2\pi)}{2\pi i} \right)\,,\qquad B=\left(\frac{23}{6},\,1,\,\frac{23}{12}\right)^T \,.
} 
Moreover, the leading instanton contributions are
\eq{ 
\mathcal{F}_{\rm inst}&=-\frac{5 i \,q_{U1}}{36 \pi^3}-\frac{493 \, q_{U^1}^2}{10368 \pi^3} + \frac{5i \,q_{U^1} q_Z}{36 \pi^3}+\ldots \\[10pt]
&= -\frac{120 i }{\pi^3} e^{2\pi i U^1}-\frac{35496 i }{ \pi^3}
e^{4\pi i U^1}+ \frac{120 }{\pi^2}e^{2\pi i U^1} Z +\cdots \,.
}
The generic relation  \eqref{Umin} provides a minimum at $U^\alpha \sim
S$ which is flat along $S$ as long as the following condition on the
fluxes is satisfied
\eq{
\label{eq:min_conditions}
\vec U&= S\begin{pmatrix}p_1\\p_2\end{pmatrix}=
S \frac{i h_2}{2 \tilde{f}^{1} + \tilde{f}^{3}}
\begin{pmatrix} -1 \\ \frac{4\tilde f^1+\tilde f^2+2\tilde f^3}{2 \tilde{f}^{1} + \tilde{f}^{3}} \end{pmatrix}\,,
\\
h_1&=\left(2+\frac{\tilde f^2}{2\tilde f^1+\tilde f^3}\right)\,h_2\,.
}
Additionally the conifold modulus is stabilized by \eqref{conifold-min} with
\eq{
M&=-2\tilde f^3
\,,\\
M_1&= f_3 -\tilde f^3\left(\frac{1}{2} + \,\frac{1-\log(2\pi)}{\pi i}\right)
\,,\\
K_1&=h_{3}-\frac{(\tilde{f}^1+\tilde{f}^3) (4 \tilde{f}^1+\tilde{f}^2+2 \tilde{f}^3)}{(2 \tilde{f}^1+\tilde{f}^3)^2}h_{2}\,.
}
Note that with the exception of $M_1$, the parameters are real and $|\zeta_0|=\frac{1}{2\pi}$ is independent of the fluxes.
Hence, the conifold modulus is guaranteed to be small for $\Re(S)\gg1$ and our trusted regimes overlap.

So to first order in $Z$, which we can trust if we can stabilize at $\Re(S)\gg1$, we have a ``perturbatively flat vacuum".
The final step is to realize a racetrack-like vacuum for $S$ with $\Re(S)\gg1$ and resulting in $|W_0|\ll1$.
The effective superpotential \eqref{Weff} evaluates to
\eq{\label{Weff_ex}
W_{\rm eff}= -\frac{5}{36 \pi^2}(2\tilde f^1+\tilde f^3)q_{U^1} -\frac{\tilde f^3}{\pi^2}q_Z \, +O({q_i}^2)  \,.
}
By now we have several constraints on the fluxes. Besides the original choices and the condition we get from the $U^\alpha$ minimization, 
we need $\Re(S)\gg1$. 
The instanton expansion is under control if $|q_i|<1$ with $q_i$ given in \eqref{qcoords}.
Altogether we have
\eq{
f_0=\tilde f^i B_i \quad & \Rightarrow \,  
\frac{2\tilde f^1+\tilde f^3}{12} \in \mathbb Z \,,
\\
h_1=\left(2+\frac{\tilde f^2}{2\tilde f^1+\tilde f^3}\right)h_2 \quad&\Rightarrow\, 
\frac{h_2 \tilde f^2}{2\tilde f^1+\tilde f^3} \in \mathbb{Z} 
}
and from the instanton series
\eq{
1>|q_{U^1}|&=\left| 864 \, \exp\left(2\pi \frac{ h_2}{2\tilde f^1+\tilde f^3} S\right) \right| \,,\\
1>|q_{U^2}|&=\left| 64\, \exp\left(2\pi \Big(\frac{4\tilde f^1+\tilde f^2+2\tilde f^3}{(2\tilde f^1+\tilde f^3)\tilde f^3}h_2 - \frac{2}{\tilde f^3}h_3 \Big)S\right) \right| \,,\\
1>|q_Z|&=\left| \frac12\, \exp\left(\pi\Big(-\frac{(\tilde f^1+\tilde f^3)(4\tilde f^1+\tilde f^2+2\tilde f^3)}{(2\tilde f^1+\tilde f^3)^2\tilde f^3}h_2 + \frac{1}{\tilde f^3}h_3\Big) S\right) \right| \,.
}
Also, it is assumed that $\tilde f^3\ne0$ and $2\tilde f^1+\tilde f^3\ne 0$ in order to be able to invert the relations of steps 1 and 2.
It is straightforward to find flux combinations that fulfill these requirements without going to very large flux numbers, e.g.
\eq{
\label{eq:fluxes_vals}
	F=\begin{pmatrix} 74\\0\\0\\0\\0\\-24\\120\\24\end{pmatrix}
\qquad
	H=\begin{pmatrix} 0 \\ -9\\3\\-4 \\ 0 \\ 0 \\0\\0\end{pmatrix}\;.
}

\begin{figure}[hbt]
    \centering
    \includegraphics[width=.8\linewidth]{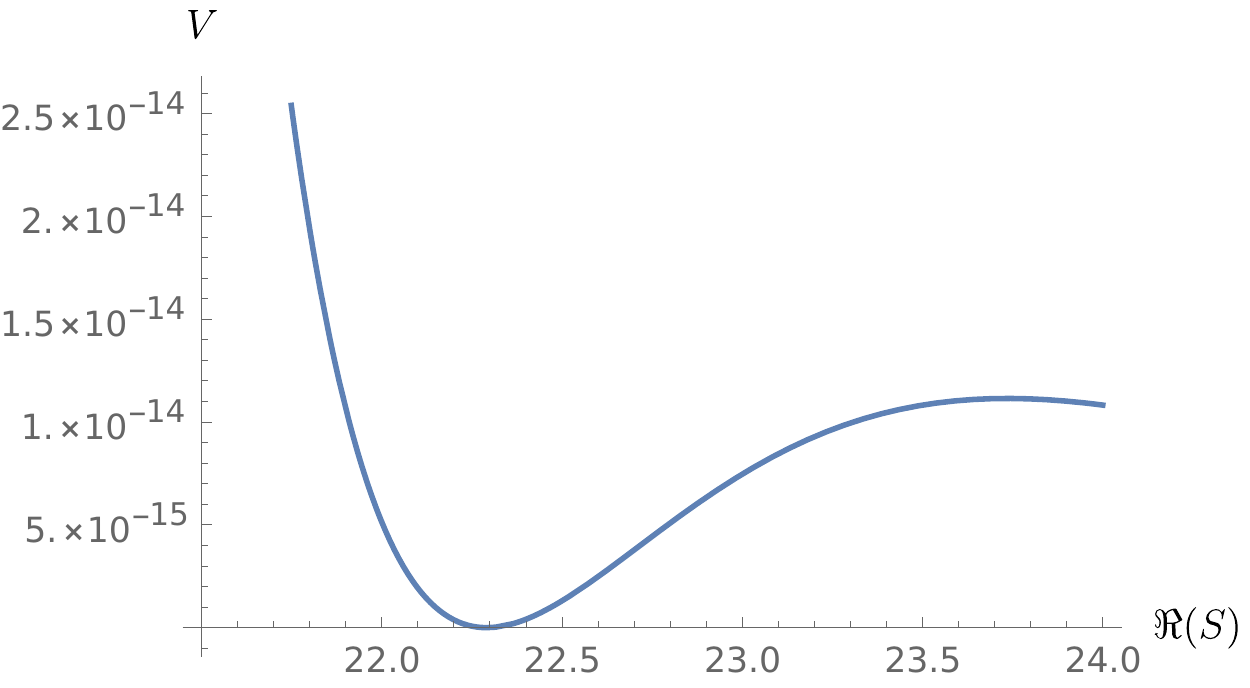}
    \caption{The effective scalar potential for the real part of $S$ shows the existence of a
Minkowski minimum.
}
    \label{fig:min}
\end{figure}

The final step is to search for a racetrack type Minkowski minimum close to the perturbatively flat minimum.
Semi-analytically minimizing the effective scalar potential for $S$, with superpotential \eqref{Weff_ex} 
evaluated along the perturbatively flat valley, we find approximate positions for the axio-dilaton (see figure \ref{fig:min}) that
lie close to the minimum of the full scalar potential depending on all
eight real scalar fields.
This true Minkowski vacuum can then be found by a numerical search using those starting points.

We have checked that in this example for the specific choice
of fluxes \eqref{eq:fluxes_vals} such a numerical minimum indeed
exists at
\eq{
    \langle U^1 \rangle = 2.79\,i,\quad
    \langle U^2 \rangle = 8.36\,i,\quad
    \langle Z \rangle = 1.36\cdot 10^{-6}i,\quad
    \langle S\rangle = 22.3\,.
}

\begin{figure}[ht]
  \begin{subfigure}[t]{.45\textwidth}
    \centering
    \includegraphics[width=\linewidth]{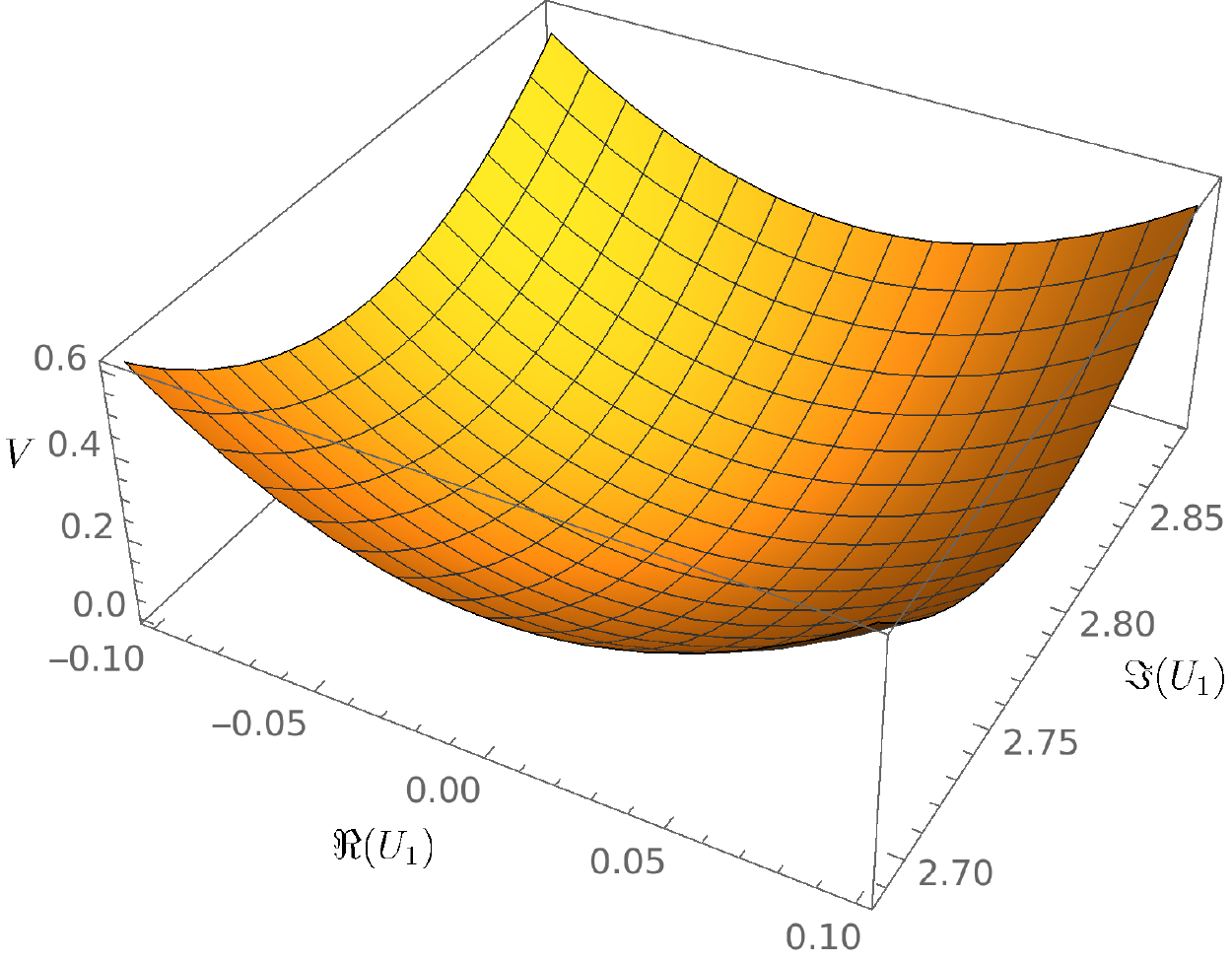}
    \caption{$V$ vs. $\Re(U^{1})$ and $\Im(U^{1})$.}
  \end{subfigure}
  \hfill
  \begin{subfigure}[t]{.45\textwidth}
    \centering
    \includegraphics[width=\linewidth]{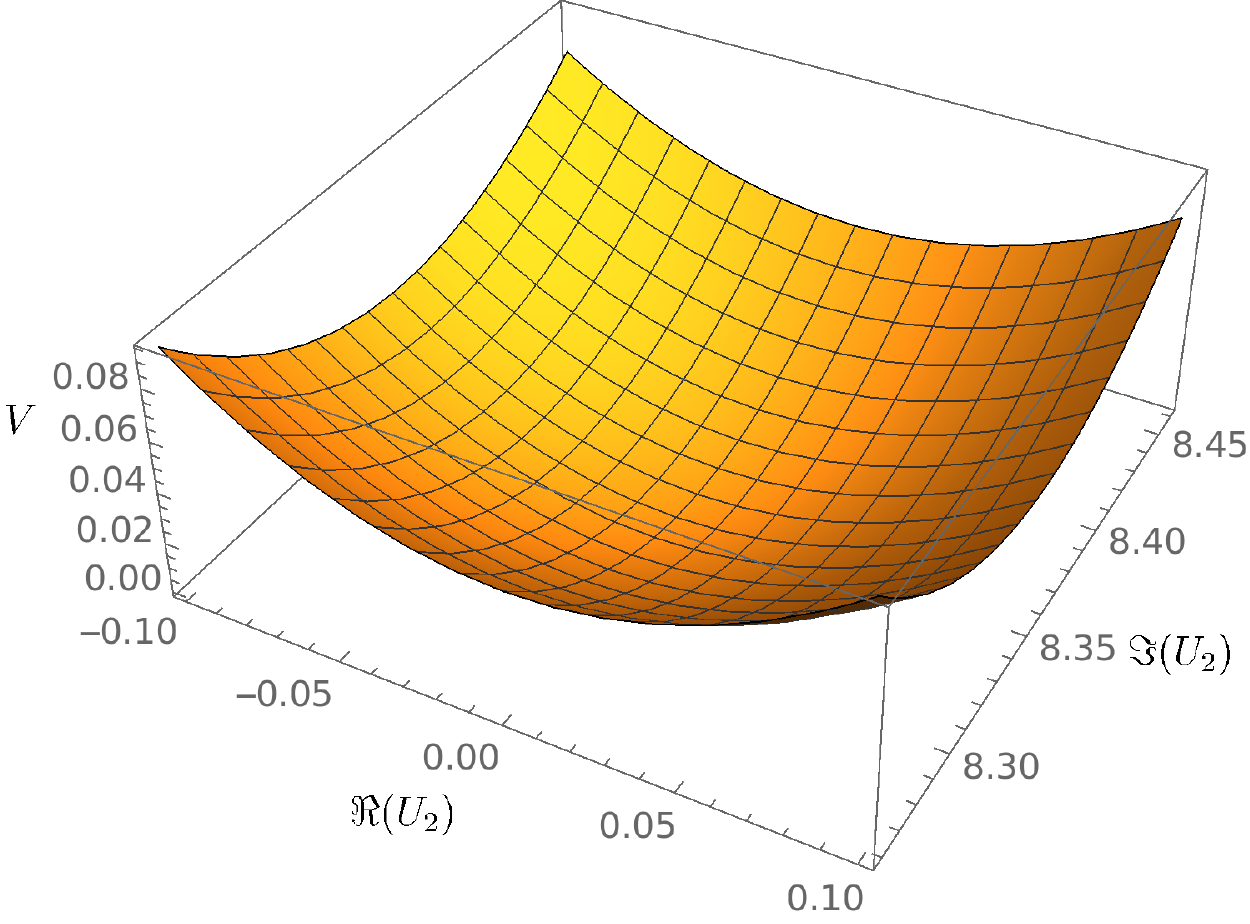}
    \caption{$V$ vs. $\Re(U^{2})$ and $\Im(U^{2})$.}
  \end{subfigure}

  \medskip

  \begin{subfigure}[t]{.45\textwidth}
    \centering
    \includegraphics[width=\linewidth]{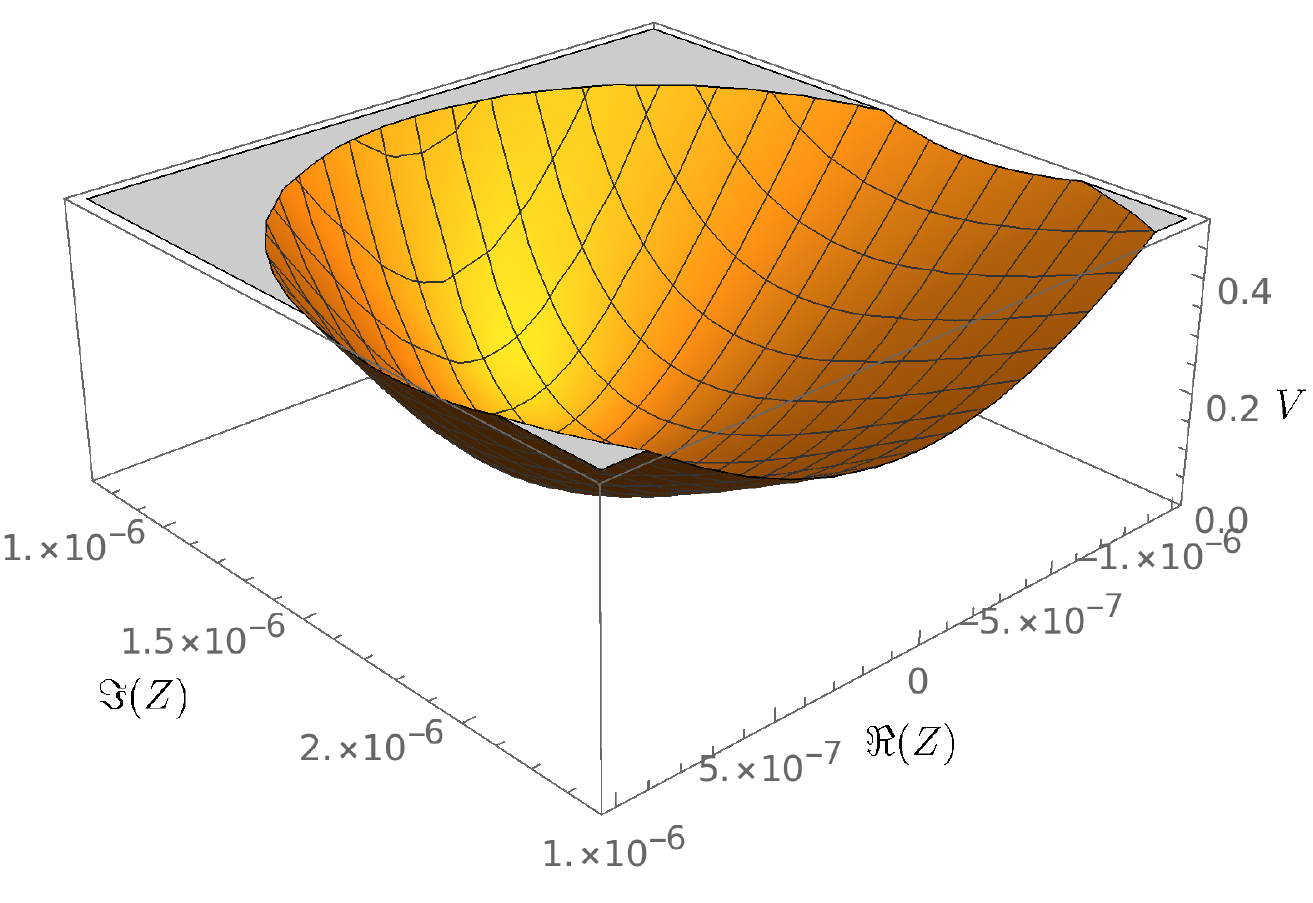}
    \caption{$V$ vs. $\Re(Z)$ and $\Im(Z)$.}
  \end{subfigure}
  \hfill
  \begin{subfigure}[t]{.45\textwidth}
    \centering
    \includegraphics[width=\linewidth]{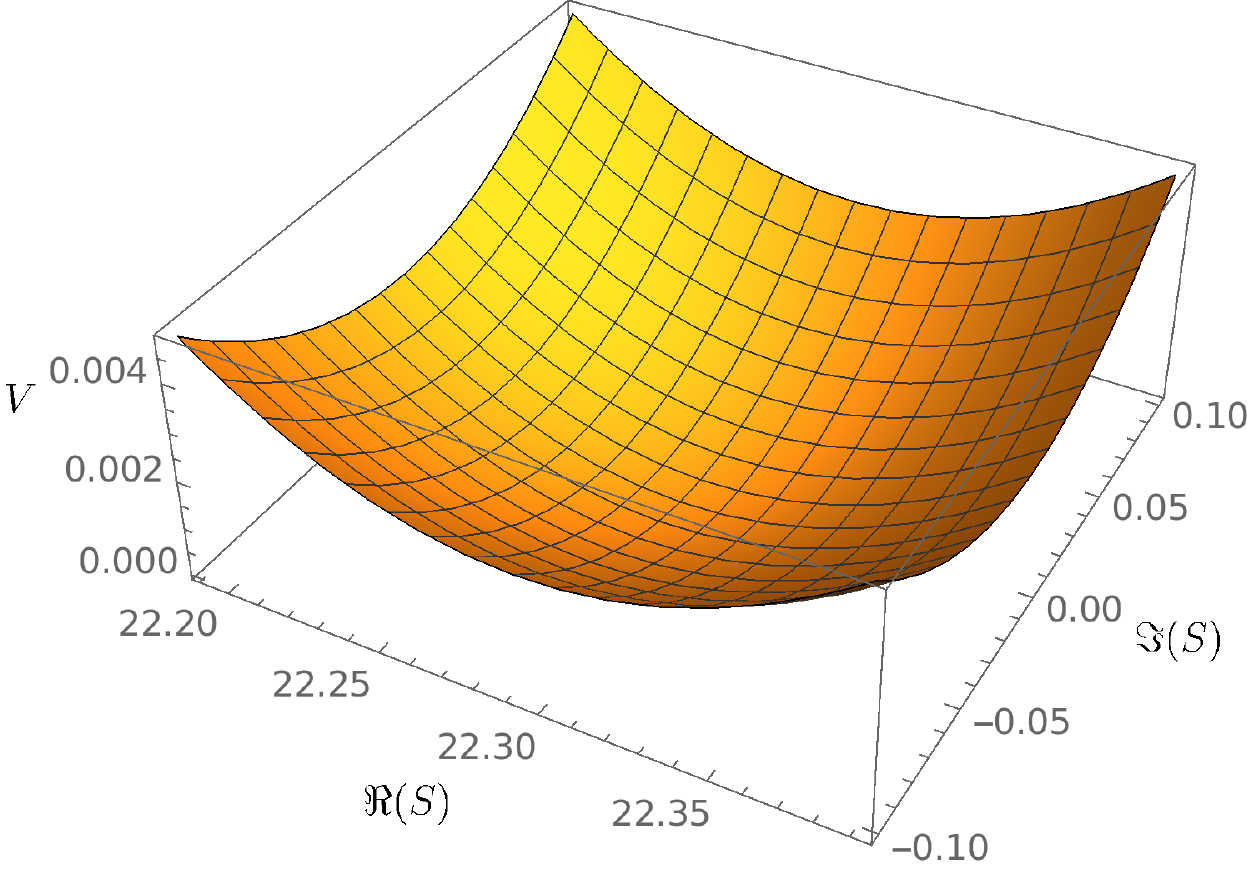}
    \caption{$V$ vs. $\Re(S)$ and $\Im(S)$.}
  \end{subfigure}  
  \caption{Full scalar potential around the minimum.\label{fig:toti}}
\end{figure}

With these values we observe that the instanton series is nicely under control with $|q_i|\approx(2\cdot10^{-5},0.2,4\cdot 10^{-6})$.
The superpotential in this minimum is very well approximated by \eqref{Weff_ex} and evaluates to 
\eq{W_0=-3.10\cdot10^{-6}\,.} Sections through the full potential are
shown in figure \ref{fig:toti}.

Computing the mass eigenvalues for our example, we find a very heavy eigenvalue corresponding to the conifold modulus, 
two less heavy directions which mix the complex structure moduli $U^i$ with the axio-dilaton, 
and a very light direction along the perturbatively flat vacuum
\eq{\{m^2\}=\{6\cdot10^{14},\,1\cdot10^3,\,3\cdot10^2,\,2\cdot10^{-11} \}M_{\rm pl}^2\,.
}
The smallest value is approximately $|W_0|^2$, which also corresponds to the mass scale of the K\"ahler modulus in the KKLT scenario. The challenge of further stabilizing the remaining moduli thus persists from the LCS point.

In an inexhaustive search over fluxes and performing the semi-analytic minimization of the effective potential for $S$ to keep the computation tractable, we find more than $10^{4}$ (approximate) vacua for which $|W_{0}|^{2} \leq 10^{-6}$, with values like $|W_0|\approx 10^{-12}$ being commonplace. 
Indeed it seems that arbitrarily small values of $W_0$ can be reached with reasonably small fluxes, 
however it is not clear if those minima are true vacua or if the approximations and numerics break down around those small values.
This has to be tested case by case using the full potential without approximations, as has been done in the example above. 
The search suggests that examples with reasonably small $W_0$ as the one discussed are nonetheless plentiful.

\section{Conclusions}
\label{sec_concl}

In this paper we have extended the construction of minima with small
values of $|W_0|$ of DKMM to the point of tangency between the
conifold and the LCS regime. We found that it is possible to construct
vacua with arbitrarily small values of $W_0$ for reasonable values
of the fluxes. As a proof of principle, the proposed construction was
successfully applied to an explicit example of  a CY 3-fold. With $O(10^2)$ fluxes we
explicitly found a minimum with $|W_0|\approx10^{-6}$, while a broad
search revealed that values of the superpotential could easily be as
small as $10^{-12}$. These examples seem to be good candidates to be
used in a KKLT-like  construction. The inclusion of the K\"ahler
moduli and their explicit stabilization \`a la KKLT was not considered
in detail. The potential issue of DKMM concerning the masses of the K\"ahler
moduli and the lightest complex structure moduli remains for future investigation.

Let us emphasize again that for the mechanism to work rational coefficients in the scalar
potential are a necessary requirement. An exact computation of these
values requires analytic knowledge of the transition matrix of the
periods to the conifold. We have shown that for a certain class of
models these can be calculated analytically using expressions for the
periods in terms of harmonic polylogarithms. Moreover, we expect this
rationality property to hold in more general models as well. The
computation in these more general cases requires evaluations or
identities between L-values of (twisted) Hecke eigenforms, which are
currently being developed \cite{Klemm1} but are beyond the scope of this 
paper. 


\vspace{0.5cm}

\noindent
\subsubsection*{Acknowledgments}
We would like to thank Albrecht Klemm for explaining  the details of
the appearance of quasi-periods to us 
as well as Mehmet Demirtas, Manki Kim, Liam McAllister and Jakob Moritz for informing
us about their upcoming work prior to submission.

\vspace{1cm}

\clearpage
\appendix

\section{Results for \texorpdfstring{$\mathbb{P}_{1,1,2,8,12}[24]$}{P(1,1,2,8,12)[24]}}

In this appendix we collect some more details about the periods of $\mathbb{P}_{1,1,2,8,12}[24]$.

\subsection{Local periods at the LCS }
\label{subsec:local_LCS_periods}

A local basis of periods $\omega_{\text{LCS}}$ around the LCS point is given by
\begin{align*}
    \omega_{\text{LCS},1} &= w_{1}\,,\\[2ex]
    \omega_{\text{LCS},2} &= w_{2} - \frac{i w_{1} \log (x_{1})}{2 \pi }\,,\\[2ex]
    \omega_{\text{LCS},3} &= w_{3} - \frac{i w_{1} \log (x_{2})}{2 \pi }\,,\\[2ex]
    \omega_{\text{LCS},4} &= w_{4} - \frac{i w_{1} \log (x_{3})}{2 \pi }\,,\\[2ex]
    \omega_{\text{LCS},5} &= w_{5} + \frac{w_{1} \log ^2(x_{1})}{\pi^2} + \frac{w_{1} \log^2(x_{3})}{4 \pi ^2} + \frac{w_{1} \log (x_{1}) \log (x_{2})}{2 \pi ^2} + \frac{w_{1} \log(x_{1}) \log (x_{3})}{\pi^2}\\
    & + \frac{w_{1} \log (x_{2})\log (x_{3})}{4\pi^2} + \left(\frac{4 i w_{2}}{\pi}+\frac{i w_{3}}{\pi}+\frac{2 i w_{4}}{\pi}\right) \log (x_{1})\\
    & + \left(\frac{2i w_{2}}{\pi }+\frac{i w_{3}}{2 \pi }+\frac{i w_{4}}{\pi }\right)\log(x_{3}) + \left(\frac{i w_{2}}{\pi}+\frac{i w_{4}}{2 \pi }\right) \log (x_{2})\,,\\[2ex]
   \omega_{\text{LCS},6} &= w_{6} + \frac{w_{1}
   \log (x_{1}) \log (x_{3})}{4 \pi ^2}+\frac{w_{1} \log ^2(x_{1})}{4 \pi
   ^2}+\left(\frac{i w_{2}}{\pi }+\frac{i w_{4}}{2 \pi }\right) \log (x_{1})+\frac{i
   w_{2} \log (x_{3})}{2 \pi }\,,\\[2ex]
   \omega_{\text{LCS},7} &= w_{7} + \frac{w_{1} \log (x_{1}) \log(x_{2})}{4 \pi ^2}+\frac{w_{1} \log (x_{1}) \log (x_{3})}{2 \pi^2}+\frac{w_{1} \log ^2(x_{1})}{2 \pi ^2}\\
   & + \left(\frac{2 i w_{2}}{\pi }+\frac{i w_{3}}{2 \pi }+\frac{i w_{4}}{\pi }\right)\log (x_{1})+\frac{i w_{2}
   \log (x_{2})}{2 \pi }+\frac{i w_{2} \log (x_{3})}{\pi }\,,\\[2ex]
   \omega_{\text{LCS},8} &= w_{8} + \frac{i w_{1} \log (x_{1}) \log (x_{2}) \log (x_{3})}{8 \pi ^3}+\frac{i w_{1} \log^2(x_{1}) \log (x_{2})}{8 \pi ^3}+\frac{i w_{1} \log ^2(x_{1}) \log
   (x_{3})}{4 \pi ^3}\\
   & + \frac{i w_{1} \log (x_{1}) \log ^2(x_{3})}{8 \pi^3}+\frac{i w_{1} \log ^3(x_{1})}{6 \pi ^3}+
   \left(-\frac{w_{2}}{\pi ^2}-\frac{w_{3}}{4 \pi ^2}-\frac{w_{4}}{2 \pi^2}\right)\log (x_{1}) \log (x_{3})\\
   &+\left(-\frac{w_{2}}{\pi ^2}-\frac{w_{3}}{4 \pi^2}-\frac{w_{4}}{2 \pi ^2}\right)\log ^2(x_{1})+\left(-\frac{w_{2}}{2 \pi ^2}-\frac{w_{4}}{4 \pi^2}\right) \log (x_{1}) \log (x_{2})\\
   &-\frac{w_{2} \log (x_{2}) \log(x_{3})}{4 \pi ^2}-\frac{w_{2} \log^2(x_{3})}{4 \pi ^2}+\frac{i w_{5} \log(x_{1})}{2 \pi }+\frac{i w_{6} \log (x_{2})}{2 \pi }+\frac{i w_{7} \log(x_{3})}{2 \pi }\,,
\end{align*}
where the power series terms are
\begin{align*}
    w_{1} &= 1+60 x_{1}+13860 x_{1}^2+27720 x_{1}^2 x_{3} + \mathcal{O}(x^{3})\,,\\[2ex]
    w_{2} &= -\frac{156 i x_{1}}{\pi }-\frac{38826 i x_{1}^2}{\pi }+\frac{i x_{3}}{2 \pi }-\frac{30 i x_{1} x_{3}}{\pi }-\frac{98442 i x_{1}^2 x_{3}}{\pi }+\frac{3 i x_{3}^2}{4 \pi }\\
    &-\frac{15 i x_{1} x_{3}^2}{\pi }+\frac{3465 i x_{1}^2 x_{3}^2}{\pi }+\frac{3 i x_{2} x_{3}^2}{2 \pi }-\frac{30 i x_{1} x_{2} x_{3}^2}{\pi }+\frac{6930 i x_{1}^2 x_{2} x_{3}^2}{\pi } + \mathcal{O}(x^{3})\,,\\[2ex]
    w_{3} &= -\frac{i x_{2}}{\pi }-\frac{60 i x_{1} x_{2}}{\pi }-\frac{13860 i x_{1}^2 x_{2}}{\pi }-\frac{3 i x_{2}^2}{2 \pi }-\frac{90 i x_{1} x_{2}^2}{\pi }-\frac{20790 i x_{1}^2 x_{2}^2}{\pi }\\
    &-\frac{27720 i x_{1}^2 x_{3}}{\pi }+\frac{27720 i x_{1}^2 x_{2} x_{3}}{\pi }+\frac{13860 i x_{1}^2 x_{2}^2 x_{3}}{\pi } + \mathcal{O}(x^{3})\,,\\[2ex]
    w_{4} &= -\frac{60 i x_{1}}{\pi }-\frac{20790 i x_{1}^2}{\pi }+\frac{i x_{2}}{2 \pi }+\frac{30 i x_{1} x_{2}}{\pi }+\frac{6930 i x_{1}^2 x_{2}}{\pi }+\frac{3 i x_{2}^2}{4 \pi }\\
    &+\frac{45 i x_{1} x_{2}^2}{\pi }+\frac{10395 i x_{1}^2 x_{2}^2}{\pi }-\frac{i x_{3}}{\pi }+\frac{60 i x_{1} x_{3}}{\pi }+\frac{27720 i x_{1}^2 x_{3}}{\pi }\\
    &-\frac{13860 i x_{1}^2 x_{2} x_{3}}{\pi }-\frac{6930 i x_{1}^2 x_{2}^2 x_{3}}{\pi }-\frac{3 i x_{3}^2}{2 \pi }+\frac{30 i x_{1} x_{3}^2}{\pi }-\frac{6930 i x_{1}^2 x_{3}^2}{\pi }\\
    &-\frac{3 i x_{2} x_{3}^2}{\pi }+\frac{60 i x_{1} x_{2} x_{3}^2}{\pi }-\frac{13860 i x_{1}^2 x_{2} x_{3}^2}{\pi } + \mathcal{O}(x^{3})\,,\\[2ex]
    w_{5} &= \frac{120 x_{1}}{\pi ^2}+\frac{183294 x_{1}^2}{\pi ^2}-\frac{x_{2}^2}{4 \pi ^2}-\frac{15 x_{1} x_{2}^2}{\pi ^2}-\frac{3465 x_{1}^2 x_{2}^2}{\pi ^2}+\frac{169704 x_{1}^2 x_{3}}{\pi ^2}\\
    &-\frac{13860 x_{1}^2 x_{2} x_{3}}{\pi ^2}  + \mathcal{O}(x^{3})\,,\\[2ex]
    w_{6} &= \frac{33696 x_{1}^2}{\pi ^2}-\frac{78 x_{1} x_{2}}{\pi ^2}-\frac{19413 x_{1}^2 x_{2}}{\pi ^2}-\frac{117 x_{1} x_{2}^2}{\pi ^2}-\frac{58239 x_{1}^2 x_{2}^2}{2 \pi ^2}\\
    &-\frac{30 x_{1} x_{3}}{\pi ^2}-\frac{6795 x_{1}^2 x_{3}}{\pi ^2}-\frac{x_{2} x_{3}}{4 \pi ^2}+\frac{15 x_{1} x_{2} x_{3}}{\pi ^2}+\frac{49221 x_{1}^2 x_{2} x_{3}}{\pi ^2}-\frac{x_{2}^2 x_{3}}{8 \pi ^2}\\
    &+\frac{15 x_{1} x_{2}^2 x_{3}}{2 \pi ^2}+\frac{49221 x_{1}^2 x_{2}^2 x_{3}}{2 \pi ^2}-\frac{x_{3}^2}{4 \pi ^2}+\frac{3465 x_{1}^2 x_{3}^2}{2 \pi ^2}-\frac{13 x_{2} x_{3}^2}{8 \pi ^2}+\frac{45 x_{1} x_{2} x_{3}^2}{2 \pi ^2}\\
    &-\frac{3465 x_{1}^2 x_{2} x_{3}^2}{2 \pi ^2}+\frac{3 x_{2}^2 x_{3}^2}{16 \pi ^2}-\frac{15 x_{1} x_{2}^2 x_{3}^2}{4 \pi ^2}+\frac{3465 x_{1}^2 x_{2}^2 x_{3}^2}{4 \pi ^2} + \mathcal{O}(x^{3})\,,\\[2ex]
    w_{7} &= \frac{67392 x_{1}^2}{\pi ^2}-\frac{x_{3}}{2 \pi ^2}-\frac{30 x_{1} x_{3}}{\pi ^2}+\frac{84852 x_{1}^2 x_{3}}{\pi ^2}-\frac{13 x_{3}^2}{8 \pi ^2}+\frac{45 x_{1} x_{3}^2}{2 \pi ^2}\\
    &-\frac{3465 x_{1}^2 x_{3}^2}{2 \pi ^2}-\frac{7 x_{2} x_{3}^2}{4 \pi ^2}+\frac{15 x_{1} x_{2} x_{3}^2}{\pi ^2}+\frac{3465 x_{1}^2 x_{2} x_{3}^2}{\pi ^2} + \mathcal{O}(x^{3})\,,\\[2ex]
    w_{8} &= -\frac{120 i x_{1}}{\pi ^3}-\frac{26055 i x_{1}^2}{\pi ^3}-\frac{39 i x_{1} x_{2}^2}{\pi ^3}-\frac{19413 i x_{1}^2 x_{2}^2}{2 \pi ^3}+\frac{i x_{3}}{2 \pi ^3}-\frac{91782 i x_{1}^2 x_{3}}{\pi ^3}\\
    &+\frac{i x_{2} x_{3}}{4 \pi ^3}-\frac{15 i x_{1} x_{2} x_{3}}{\pi ^3}-\frac{49221 i x_{1}^2 x_{2} x_{3}}{\pi ^3}+\frac{9 i x_{3}^2}{16 \pi ^3}-\frac{75 i x_{1} x_{3}^2}{4 \pi ^3}+\frac{10395 i x_{1}^2 x_{3}^2}{2 \pi ^3}\\
    &+\frac{23 i x_{2} x_{3}^2}{16 \pi ^3}-\frac{135 i x_{1} x_{2} x_{3}^2}{4 \pi ^3}+\frac{24255 i x_{1}^2 x_{2} x_{3}^2}{4 \pi ^3}-\frac{3 i x_{2}^2 x_{3}^2}{32 \pi ^3}+\frac{15 i x_{1} x_{2}^2 x_{3}^2}{8 \pi ^3}\\
    &-\frac{3465 i x_{1}^2 x_{2}^2 x_{3}^2}{8 \pi ^3} + \mathcal{O}(x^{3})\,.
\end{align*}

\subsection{Local periods at the conifold }
\label{subsec:local_coni_periods}

A local basis of periods $\omega_{\text{c}}$ around the $(\overline{x},\overline{y},\overline{z}) = (0,0,1)$ conifold is given by
\begin{align*}
    \omega_{\text{c},1} &= \tilde{w}_{1}\,,\\[2ex]
    \omega_{\text{c},2} &= \tilde{w}_{2} + \tilde{w}_{1} \log (x_{1})\,,\\[2ex]
    \omega_{\text{c},3} &= \tilde{w}_{3} + \frac{1}{2} \tilde{w}_{1} \log (x_{2})+\tilde{w}_{1} \log (x_{3})\,,\\[2ex]
    \omega_{\text{c},4} &= \tilde{w}_{4}\,,\\[2ex]
    \omega_{\text{c},5} &= \tilde{w}_{5} + \tilde{w}_{4} \log (x_{2})\,,\\[2ex]
    \omega_{\text{c},6} &= \tilde{w}_{6} + \tilde{w}_{1} \log ^2(x_{1})+2 \tilde{w}_{2} \log (x_{1})\,,\\[2ex]
    \omega_{\text{c},7} &= \tilde{w}_{7} + \frac{1}{2} \tilde{w}_{1} \log (x_{1}) \log (x_{2})+\tilde{w}_{1} \log (x_{1}) \log (x_{3})+\tilde{w}_{1} \log ^2(x_{1})\\
    & + (2 \tilde{w}_{2}+\tilde{w}_{3}) \log (x_{1})+\frac{1}{2} \tilde{w}_{2} \log (x_{2})+\tilde{w}_{2} \log (x_{3})\,,\\[2ex]
    \omega_{\text{c},8} &= \tilde{w}_{8} + \frac{3}{4} \tilde{w}_{1} \log ^2(x_{1}) \log (x_{2})+\frac{3}{2} \tilde{w}_{1} \log ^2(x_{1}) \log (x_{3})+\tilde{w}_{1} \log ^3(x_{1})\\
    & + \left(3 \tilde{w}_{2}+\frac{3 \tilde{w}_{3}}{2}\right) \log ^2(x_{1})+\frac{3}{2} \tilde{w}_{2} \log (x_{1}) \log (x_{2})+3 \tilde{w}_{2} \log (x_{1}) \log (x_{3})\\
    & + \frac{3}{4} \tilde{w}_{6} \log (x_{2})+\frac{3}{2} \tilde{w}_{6} \log (x_{3})+3 \tilde{w}_{7} \log (x_{1})\,,
\end{align*}
where the power series terms are
\begin{align*}
    \tilde{w}_{1} &= 1+\frac{5 x_{1}}{36}+\frac{385 x_{1}^2}{3456}-\frac{385 x_{1}^2 x_{3}}{10368} + \mathcal{O}(x^{3})\,,\\[2ex]
    \tilde{w}_{2} &= \frac{31 x_{1}}{36}+\frac{15637 x_{1}^2}{20736}-\frac{x_{3}}{2}-\frac{5 x_{1} x_{3}}{72}-\frac{2927 x_{1}^2 x_{3}}{10368}-\frac{x_{3}^2}{4}-\frac{5 x_{1} x_{3}^2}{144}-\frac{385 x_{1}^2 x_{3}^2}{41472}\\
    & - \frac{x_{2} x_{3}^2}{8}-\frac{5}{288} x_{1} x_{2} x_{3}^2-\frac{385 x_{1}^2 x_{2} x_{3}^2}{82944} + \mathcal{O}(x^{3})\,,\\[2ex]
    \tilde{w}_{3} &= \frac{385 x_{1}^2}{10368}+x_{3}+\frac{5 x_{1} x_{3}}{36}+\frac{385 x_{1}^2 x_{3}}{5184}+\frac{x_{3}^2}{2}+\frac{5 x_{1} x_{3}^2}{72}+\frac{385 x_{1}^2 x_{3}^2}{20736}+\frac{x_{2} x_{3}^2}{4}\\
    & + \frac{5}{144} x_{1} x_{2} x_{3}^2+\frac{385 x_{1}^2 x_{2} x_{3}^2}{41472} + \mathcal{O}(x^{3})\,,\\[2ex]
    \tilde{w}_{4} &= \sqrt{x_{3}}-\frac{x_{2} \sqrt{x_{3}}}{16}-\frac{15 x_{2}^2 \sqrt{x_{3}}}{1024}+\frac{x_{3}^{3/2}}{3}+\frac{5}{108} x_{1} x_{3}^{3/2}+\frac{1}{16} x_{2} x_{3}^{3/2}+\frac{5}{576} x_{1} x_{2} x_{3}^{3/2}\\
    & + \frac{3 x_{2}^2 x_{3}^{3/2}}{1024}+\frac{5 x_{1} x_{2}^2 x_{3}^{3/2}}{12288} + \mathcal{O}(x^{3})\,,\\[2ex]
    \tilde{w}_{5} &= 2 \sqrt{x_{3}}-\frac{17 x_{2}^2 \sqrt{x_{3}}}{1024}+\frac{10 x_{3}^{3/2}}{9}+\frac{25}{162} x_{1} x_{3}^{3/2}-\frac{1}{4} x_{2} x_{3}^{3/2}-\frac{5}{144} x_{1} x_{2} x_{3}^{3/2}+\frac{x_{2}^2 x_{3}^{3/2}}{1024}\\
    & + \frac{5 x_{1} x_{2}^2 x_{3}^{3/2}}{36864} + \mathcal{O}(x^{3})\,,\\[2ex]
    \tilde{w}_{6} &= \frac{961 x_{1}^2}{1296}-x_{3}-\frac{13 x_{1} x_{3}}{18}-\frac{13727 x_{1}^2 x_{3}}{20736}-\frac{5 x_{3}^2}{12}-\frac{19 x_{1} x_{3}^2}{48}-\frac{443 x_{1}^2 x_{3}^2}{5184}-\frac{5 x_{2} x_{3}^2}{24}\\
    & - \frac{19}{96} x_{1} x_{2} x_{3}^2-\frac{443 x_{1}^2 x_{2} x_{3}^2}{10368} + \mathcal{O}(x^{3})\,,\\[2ex]
    \tilde{w}_{7} &= \frac{5 x_{1}}{18}+\frac{1045 x_{1}^2}{864}-\frac{5099 x_{1}^2 x_{3}}{20736}-\frac{x_{3}^2}{4}-\frac{5 x_{1} x_{3}^2}{144}-\frac{385 x_{1}^2 x_{3}^2}{20736}-\frac{385 x_{1}^2 x_{2} x_{3}^2}{82944} + \mathcal{O}(x^{3})\,,\\[2ex]
    \tilde{w}_{8} &= -\frac{5 x_{1}}{3}-\frac{1313 x_{1}^2}{864}+3 x_{3}+\frac{2557 x_{1}^2 x_{3}}{6912}+\frac{2 x_{3}^2}{3}-\frac{7 x_{1} x_{3}^2}{16}-\frac{719 x_{1}^2 x_{3}^2}{2304}+\frac{31 x_{2} x_{3}^2}{48}\\
    & + \frac{5}{64} x_{1} x_{2} x_{3}^2-\frac{1271 x_{1}^2 x_{2} x_{3}^2}{13824} + \mathcal{O}(x^{3})\,.\\[2ex]
\end{align*}

\subsection{Symplectic form }
\label{subsec:symplectic_form_coefficients}

The coefficients of the symplectic form are
\begingroup
\allowdisplaybreaks
\begin{align*}
    A_{1}(\overline{x},\overline{y},\overline{z}) &= -\frac{\overline{x} \overline{y} \overline{z}^2 \left(62 \overline{x}^3 (\overline{z}-1)+10 \overline{x}^2 (\overline{z}-2)+15 \overline{x}-5\right)}{288 (\overline{x}-1) (\overline{z}-1) \left(\overline{x}^2 (\overline{z}-1)+2 \overline{x}-1\right)}\eta\,,\\
    A_{2}(\overline{x},\overline{y},\overline{z}) &= -\frac{5 \overline{x}^2 (\overline{y}-1) \overline{z}}{36 (\overline{x}-1)}\eta\,,\\
    A_{3}(\overline{x},\overline{y},\overline{z}) &= -\frac{5 \overline{x} \overline{y} \overline{z} \left(2 \overline{x}^3 (\overline{z}-1)-2 \overline{x}^2 \left(\overline{z}^2+\overline{z}-2\right)-2 \overline{x}+\overline{z}\right)}{144 (\overline{x}-1) (\overline{z}-1)
   \left(\overline{x}^2 (\overline{z}-1)+2 \overline{x}-1\right)}\eta\,,\\
    A_{4}(\overline{x},\overline{y},\overline{z}) &= -\frac{\overline{x}^2 (\overline{y}-1) \overline{z}}{2 (\overline{x}-1)}\eta\,,\\
    A_{5}(\overline{x},\overline{y},\overline{z}) &= -\frac{\overline{x}^2 \overline{y} \overline{z} \left(\overline{x}^2 (2 \overline{z}+1)-2 \overline{x}+1\right)}{4 (\overline{x}-1) \left(\overline{x}^2 (\overline{z}-1)+2 \overline{x}-1\right)}\eta\,,\\
    A_{6}(\overline{x},\overline{y},\overline{z}) &= 0\,,\\
    A_{7}(\overline{x},\overline{y},\overline{z}) &= -\frac{\overline{x}^2 (\overline{y}-1) \overline{z}}{2 (\overline{x}-1)}\eta\,,\\
    A_{8}(\overline{x},\overline{y},\overline{z}) &= \frac{\overline{x}^2 \overline{y} \overline{z} (2 \overline{z}-1)}{4 (\overline{x}-1) (\overline{z}-1)}\eta\,,\\
    A_{9}(\overline{x},\overline{y},\overline{z}) &= 0\,,\\
    A_{10}(\overline{x},\overline{y},\overline{z}) &= -\frac{1}{2}  (2 \overline{x}-1) (\overline{y}-1) \overline{z}\eta\,,\\
    A_{11}(\overline{x},\overline{y},\overline{z}) &= -\frac{(2 \overline{x}-1) \overline{y} \overline{z}}{4 (\overline{z}-1)}\eta\,,\\
    A_{12}(\overline{x},\overline{y},\overline{z}) &= \frac{(\overline{x}-1) \left((\overline{y}-1) \overline{z}^2+2 \overline{z}-1\right)}{\overline{z}-1}\eta\,,\\
    A_{13}(\overline{x},\overline{y},\overline{z}) &= 2  (\overline{x}-1) (\overline{y}-1)\eta\,,\\
    A_{14}(\overline{x},\overline{y},\overline{z}) &= \frac{(\overline{y}-1) \left(\overline{x}^2 (\overline{z}-1)+2 \overline{x}-1\right)}{\overline{x}-1}\eta\,,\\
    A_{15}(\overline{x},\overline{y},\overline{z}) &= 0\,,\\
    A_{16}(\overline{x},\overline{y},\overline{z}) &= (\overline{x}-1) (\overline{y}-1) (\overline{z}-1)\eta\,,\\
    A_{17}(\overline{x},\overline{y},\overline{z}) &= -\frac{\overline{y} (2 \overline{z}-1) \left(\overline{x}^2 (\overline{z}-1)+2 \overline{x}-1\right)}{2 (\overline{x}-1) (\overline{z}-1)}\eta\,,\\
    A_{18}(\overline{x},\overline{y},\overline{z}) &= \frac{2 \left((\overline{y}-1) \overline{z}^2+2 \overline{z}-1\right)}{\overline{z}-1}\eta\,,\\
    A_{19}(\overline{x},\overline{y},\overline{z}) &= -\frac{\left(\overline{x}^4 \left((\overline{y}-1) \overline{z}^2+2 \overline{z}-1\right)-4 \overline{x}^3 (\overline{z}-1)+2 \overline{x}^2 (\overline{z}-3)+4 \overline{x}-1\right)}{(\overline{x}-1) \left(\overline{x}^2 (\overline{z}-1)+2 \overline{x}-1\right)}\eta\,.
\end{align*}
\endgroup

\section{Definitions}
\subsection{Harmonic Polylogarithms}\label{polylog}
In this section we give the basic definitions of the used harmonic polylogarithms (HPL). These as well as a Mathematica package to evaluate them can be found in \cite{Maitre:2005uu}.
HPLs are one-variable functions with a parameter vector $\vec{a}$. The dimension $k$ of the vector $a$ is called the weight of the HPL. We define the functions 
\begin{eqnarray}
f_1(x)&=&\frac{1}{1-x}\nonumber\\
f_0(x)&=&\frac{1}{x}\nonumber\\
f_{-1}(x)&=&\frac{1}{1+x}
\end{eqnarray}
The HPL's are defined recursively through integration of these three functions:
\begin{equation}
\text{HPL}(a,a_1,\dots,a_k;x)=\int\limits_0^x f_a(t)\,
\text{HPL}(a_1,\dots,a_k;t)\, \text{d} t\;. 
\end{equation}
For the weight one $\text{HPL}\left[-1;\frac{1-x_3}{1+x_3}\right]$ from the main text we have 
\begin{eqnarray}
\text{HPL}\left[-1;\frac{1-x_3}{1+x_3}\right]=\int\limits_0^\frac{1-x_3}{1+x_3}\frac{1}{1+t}\text{dt}=\log\left(1+\frac{1-x_3}{1+x_3}\right).
\end{eqnarray}

\subsection{L-functions and Hecke operators}\label{lfunction}
In this section we give a brief definition of critical L-function values. For more details we refer to the literature.
Given the $q$-series expansion of a weight $k$ modular function $f$

\begin{equation}
    f(\tau)=\sum_{n\ge 0} a_n \, q^n\;,
\end{equation}
where $q=e^{2\pi i \tau}$, its corresponding L-function is defined as

\begin{equation}
    L(f,x)=\sum_{n\ge 0} \frac{a_n}{n^x}\;.
\end{equation}
A value $L(f,j)$ is called a critical L-value if $j\in \{1,2,\dots,k-1\}$.
The Hecke operators $T_m$ are defined by their action on a modular form as
\begin{equation}
    T_m f(\tau)= m^{k-1}\sum_{d|m}d^{-k}\sum\limits_{b=0}^{d-1}f\left(\frac{m\tau+bd}{d^2}\right)\;.
\end{equation}
A modular form which is an eigenfunction of all Hecke operators is called a Hecke eigenform, i.e.
\begin{equation}
    T_m f(\tau)=\lambda_mf(\tau)\;.
\end{equation}


\clearpage
\bibliography{references}  
\bibliographystyle{utphys}


\end{document}